\documentclass[12pt]{iopart}
\usepackage{graphicx,amssymb,amsbsy}
\begin{document}

\title{Zipf's law unzipped}
\author{Seung Ki Baek$^1$, Sebastian Bernhardsson$^2$, Petter Minnhagen$^1$}
\address{$^1$Integrated Science Laboratory, Department of Physics, Ume{\aa}
University, 901 87 Ume{\aa}, Sweden}
\address{$^2$Center for Models of Life, Niels Bohr Institute, Blegdamsvej 17
DK-2100 Copenhagen $\O$, Denmark}
\ead{Petter.Minnhagen@physics.umu.se}

\begin{abstract}
Why does Zipf's law give a good description of data from
seemingly completely unrelated phenomena? Here it is argued that the reason
is that they can all be described as outcomes of a ubiquitous random group
division: the elements can be citizens of a country and the groups family
names, or the elements can be all the words making up a novel and the groups
the unique words, or the elements could be inhabitants and the groups the
cities in a country, and so on. A Random Group Formation (RGF) is presented
from which a Bayesian estimate is obtained based on minimal information: it
provides the best prediction for the number of groups with $k$ elements,
given the total number of elements, groups, and the number of elements in
the largest group. For each specification of these three values, the RGF
predicts a unique group distribution $N(k)\propto \exp(-bk)/k^{\gamma}$,
where the power-law index $\gamma$ is a unique function of the same three
values. The universality of the result is made possible by the fact that no
system specific assumptions are made about the mechanism responsible for the
group division. The direct relation between $\gamma$ and the total number of
elements, groups, and the number of elements in the largest group, is
calculated. The predictive power of the RGF model is demonstrated by direct
comparison with data from a variety of systems. It is shown that $\gamma$
usually takes values in the interval $1\leq\gamma\leq 2$ and that the value
for a given phenomena depends in a systematic way on the total size of the
data set. The results are put in the context of earlier discussions on
Zipf's and Gibrat's laws, $N(k)\propto k^{-2}$ and the connection between
growth models and RGF is elucidated.
\end{abstract}



\maketitle

\section{Introduction}

The remarkable feature that power-law distributions are commonly
encountered in a huge variety of seemingly very different systems has a long
history: the first discovery seems to go back to Pareto's paper from 1896
concerning the uneven distribution of incomes~\cite{pareto96}. Some twenty
years later Auerbach in Ref.~\cite{auerbach13} found the same power-law
distributions for city sizes. Subsequently, George Kingsley Zipf found
power-law distributions for the word frequency in written texts and this
empirical finding became known as Zipf's law~\cite{Zipf32,Zipf35,Zipf49},
although the first discovery for the case of word frequencies was made some
twenty years earlier by J.~K. Estroup~\cite{estroup16}. By now the literature
related to Zipf's law is immense and spans basically all fields: economy,
sociology, linguistics, physics, mathematical statistics, to mention a few.
A short history can be found in Ref.~\cite{mitzenmacher03}.

A large amount of the literature on Zipf's law is concerned with empirically
finding systems which obey Zipf's law~\cite{newman05}, the precise
mathematical form of the distribution which should be associated with Zipf's
empirical law~\cite{clauset09}, statistical methods for establishing whether
or not one mathematical form fits the empirical data better than another,
and last but not least various methods for data analysis in order to find
the precise value of the power-law exponent for the alleged power law. This
is \emph{not} a concern of the present paper. Instead we focus on the
question ``Why?'': Why does Zipf's law give a good description of data from
seemingly completely unrelated phenomena?

The present work stems from a specific remark made by Herbert Simon in
Ref.~\cite{simon55} `\emph{No one supposes that there is any connection
between horse-kicks suffered by soldiers in the German army and blood cells
on a microscopic slide other than that the same urn scheme provides a
satisfactory abstract model for both phenomena}.' This is precisely the
view taken in the present paper: if a vast amount of seemingly unrelated
phenomena share a common characteristic, this characteristic cannot depend on
the details of the system but must be traced to a global feature. Simon's
explicit attempt to find such an abstract model, the Simon growth model, was
criticized by Benoit Mandelbrot in Ref.~\cite{mandelbrot59}, who instead
argued that a growth model was not adequate and that the common feature
should be associated with information and entropy~\cite{mandelbrot53}. These
opposite view-points led to a heated argument between Simon and
Mandelbrot~\cite{mitzenmacher03}. From our perspective, both were right: The
common element must be a global abstract model and information is the shared
quantity which decides the characteristics.

The basic element proposed here is that the abstract common feature is the
division into groups. The generic model for this is numbered balls divided
into boxes. As examples, we take people (balls) divided into cities (boxes),
people (balls) divided into family names (boxes), and words from a novel
(balls) divided into number of occurrences in the text (boxes). We then use
information theory to obtain the best (Bayesian) prediction of the box-size
distribution based on maximum mutual information.

In \sref{sec:what}, we present the empirical data to which we compare our
predictions. In the following \sref{sec:rgf}, we describe and explain the
Random Group Formation (RGF) model. In \sref{sec:predict}, the empirical data
is directly compared with the explicit predictions of the RGF model. The
reason for a systematic size change of the power-law exponent is explained
and exemplified by data from novels. In \sref{sec:growth}, the connection
between the equilibrium maximum-entropy distribution and Gibrat's growth
model~\cite{gibrat30,gibrat31} is discussed and it is explained that the
power-law distribution $P(k)\propto k^{-2}$ is indeed an equilibrium feature
rather than a growth feature. Finally, \sref{sec:summary} contains a summary
and concluding remarks. In addition, various more detailed clarifications are
relegated to three appendices.

\section{What do we want to explain?}
\label{sec:what}

\begin{table}
\caption{\label{table:data} Basic quantities of the datasets. $M$ is the
total number of elements, $N$ is the number of groups, $k_{\max}$ is the
size of the largest group, and $k_0$ is the size of the smallest group shown
in the dataset.}
\begin{indented}
\item[] \begin{tabular}{@{}rrrrr}
\br
            & $M$ & $N$ & $k_{\max}$ & $k_0$ \\
\mr
US counties & 277~537~173 & 2~445 & 9~519~338 & $10^4$ \\
French communes & 51~107~816 & 9~011 & 852~395 & $10^3$ \\
US family names & 242~121~073 & 151~671 & 2~376~206 & $10^2$\\
Korean family names & 45~974~571 & 244 & 9~925~949 & $10^2$\\
Hardy & 1~342~258 & 30~744 & 74~165 & $10^0$ \\
Melville & 743~666 & 30~122 & 49~136 & $10^0$ \\
\br
\end{tabular}
\end{indented}
\end{table}

\begin{figure}
\begin{center}
\includegraphics[width=0.30\textwidth]{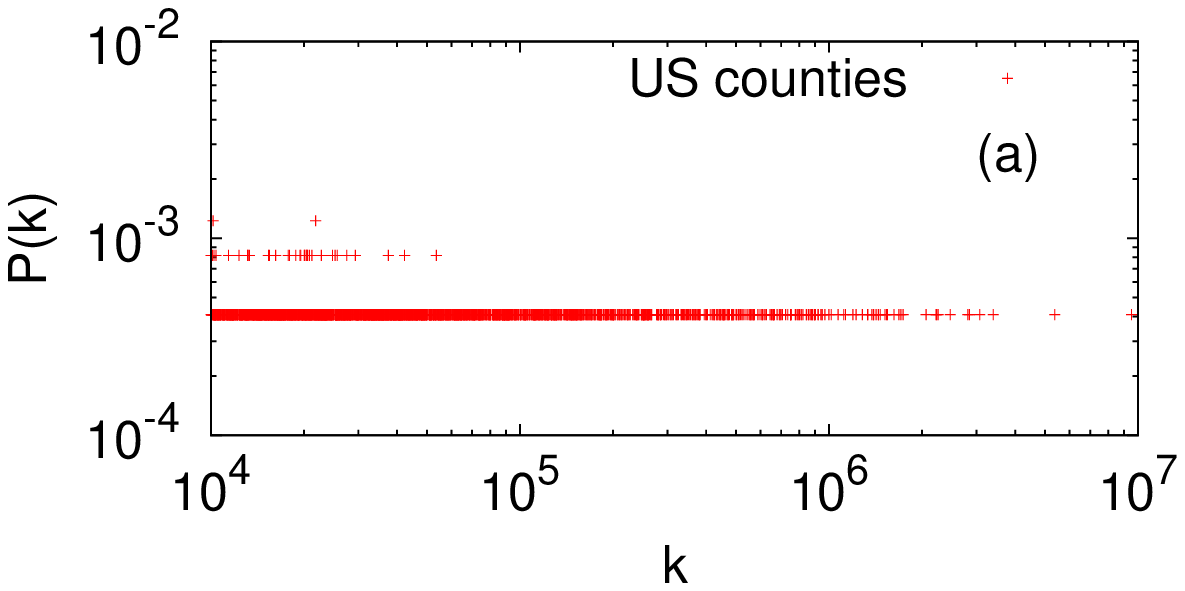}
\includegraphics[width=0.30\textwidth]{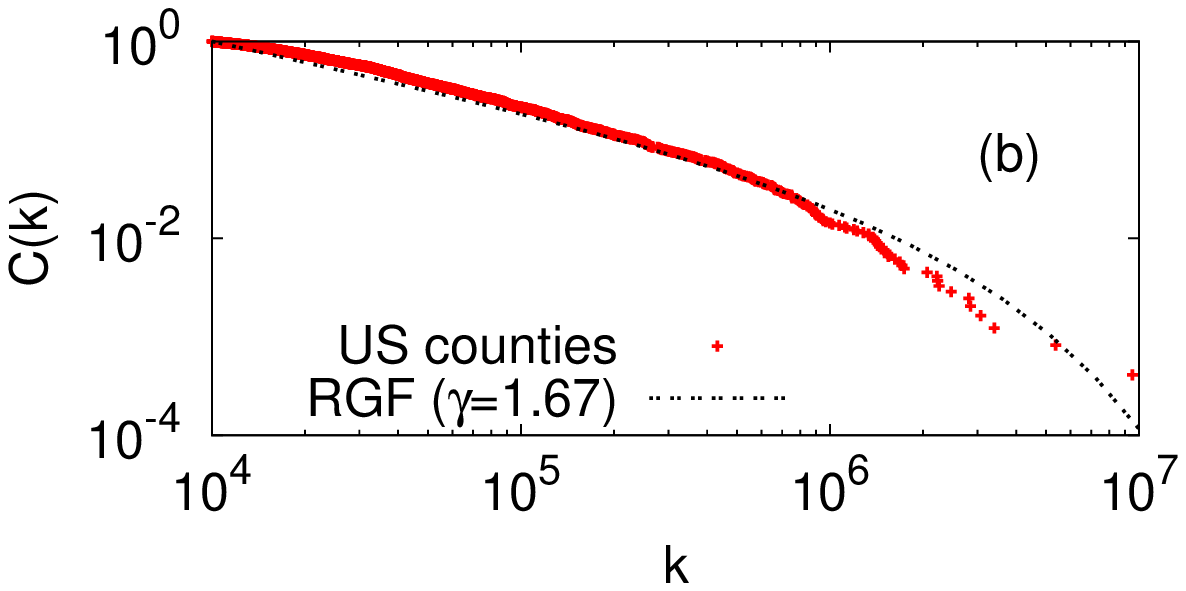}
\includegraphics[width=0.30\textwidth]{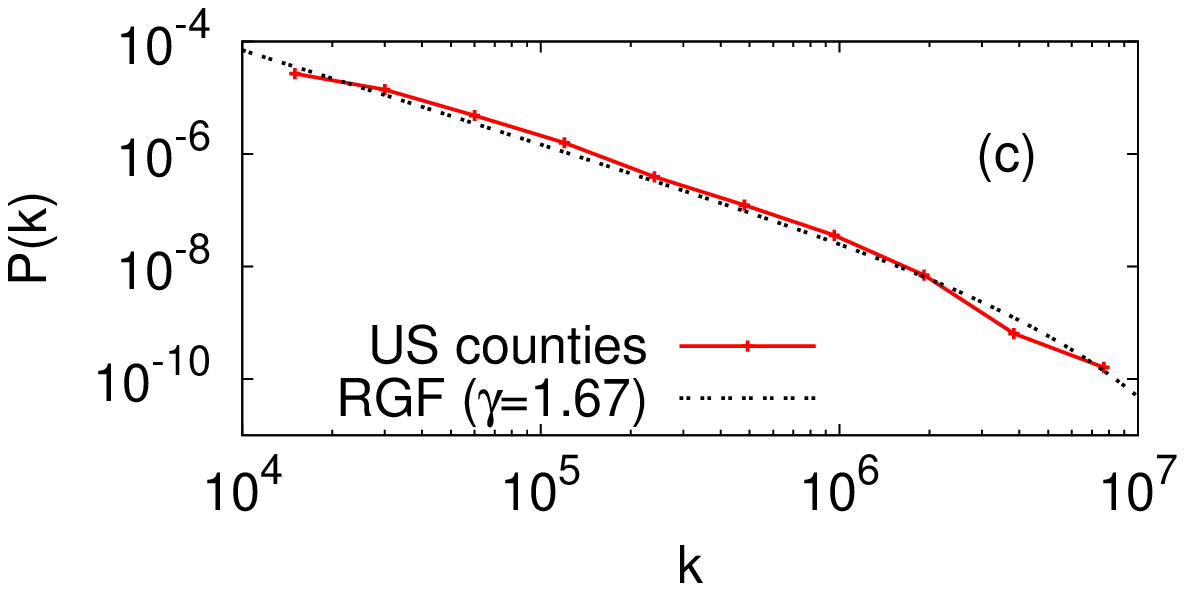}
\end{center}
\caption{Population distribution for the counties in US from the year 2000
survey (source: US Census 2000~\cite{county}).
(a) Raw-data in a log-log plot; (b) Cumulative C(k); (c) Binned P(k). Dashed
curves in (b) and (c) give the RGF prediction for the dataset.
}
\label{fig:county1}
\end{figure}

\begin{figure}
\begin{center}
\includegraphics[width=0.30\textwidth]{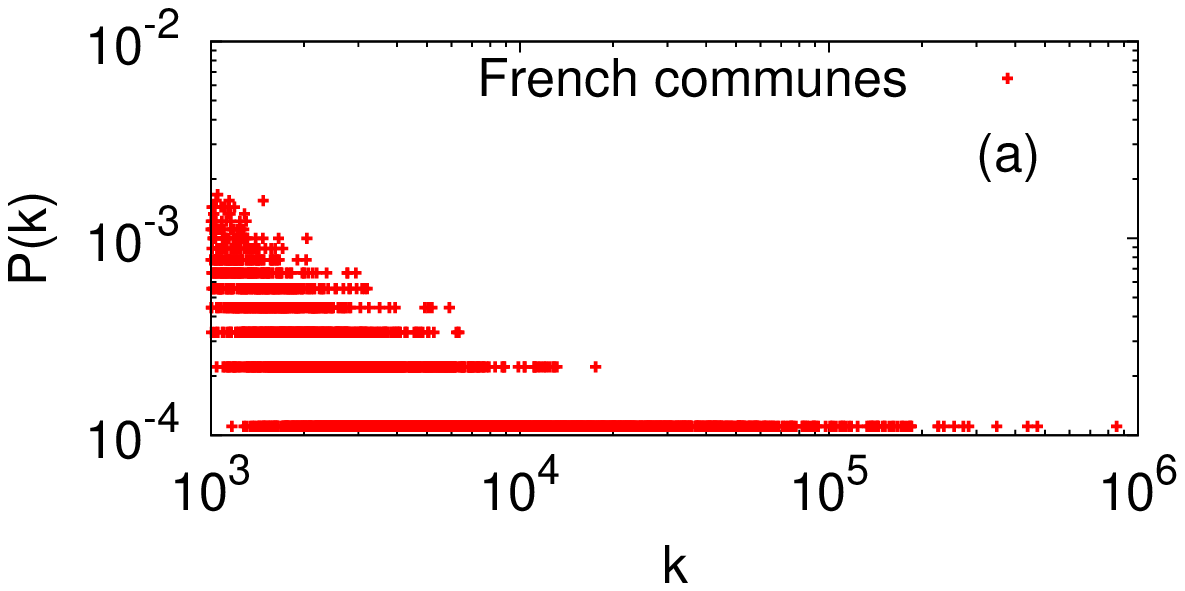}
\includegraphics[width=0.30\textwidth]{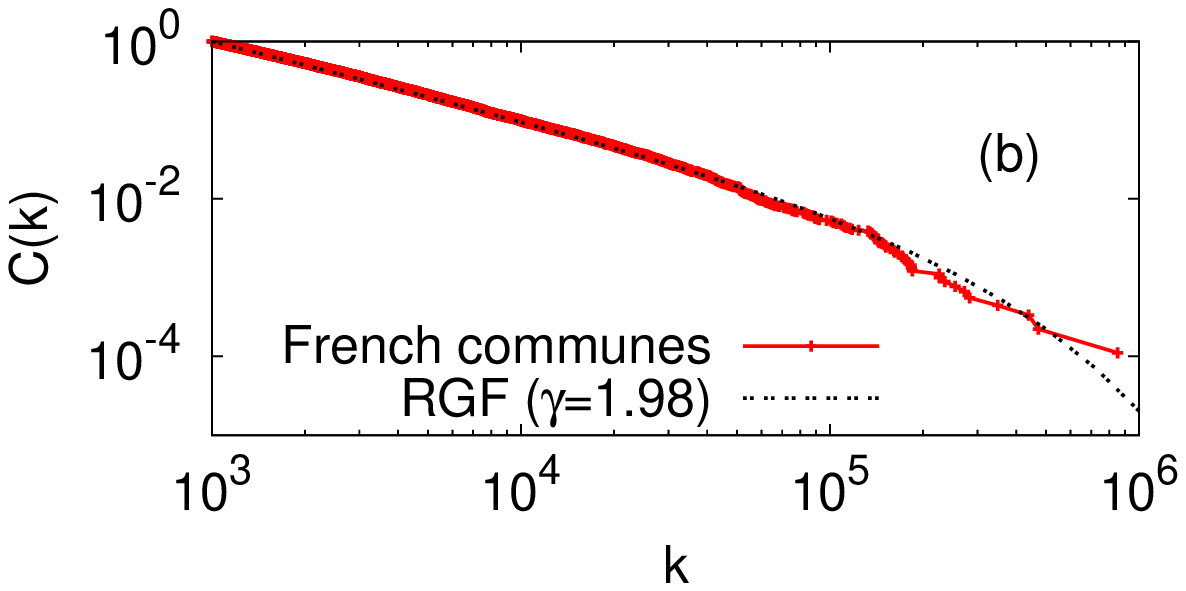}
\includegraphics[width=0.30\textwidth]{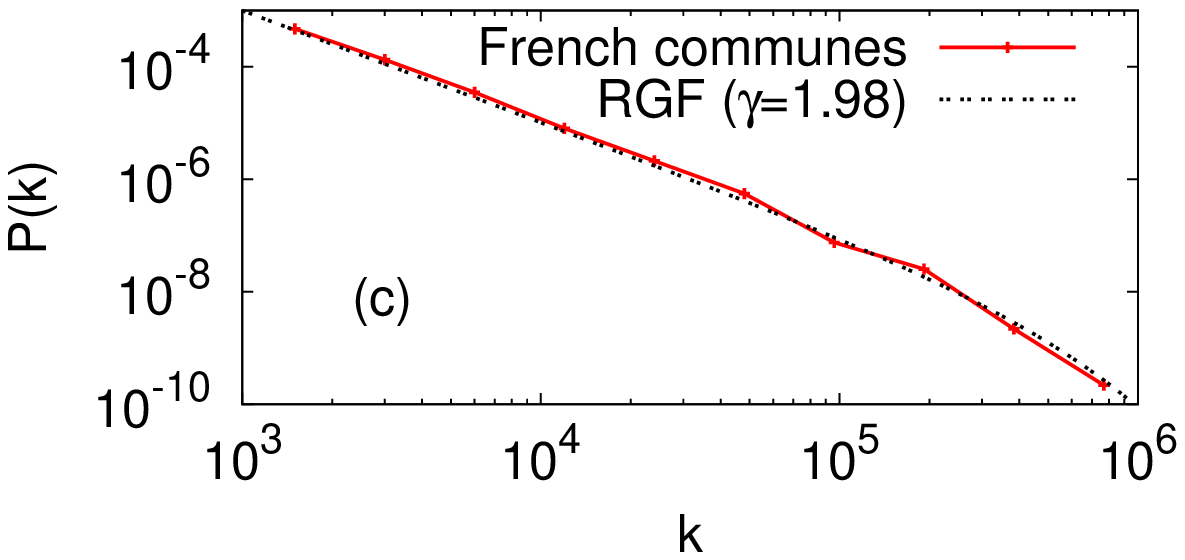}
\end{center}
\caption{ Distribution of population of French communes
(source: City Population~\cite{commune}).
(a) Raw-data in a log-log plot; (b) Cumulative C(k); (c) Binned P(k). Dashed
curves in (b) and (c) give the RGF prediction for the dataset.
} 
\label{fig:france1}
\end{figure}

The question we want to address is best illustrated by explicit examples.
Three seemingly completely unrelated phenomena are chosen: the city-size
distribution of a country, family-name frequencies for a country, and the
word-frequency distribution in novels. Two examples are given in each case.
\Fref{fig:county1} shows the county-size distribution in United States (US)
for year 2000~\cite{county} and in this case the total population is
$M=2.8\times 10^8$,
the number of counties $N=2445$ and the largest county is Los Angeles
with $k_{\max}=9.5\times 10^6$ inhabitants (see \tref{table:data}).
\Fref{fig:county1}(a)
gives the average number of counties having $k$ inhabitants and since only
rarely two counties have precisely the same number, basically all data
points fall on the line $1/N$. However, smaller counties are much more
common than very large and in \fref{fig:county1}(b) this feature is clearly
displayed by instead plotting the number of counties which have population
larger than $k$. This is usually called the cumulative distribution,
denoted by $C(k)$, and normalized such that $C(k_0)=1$, where $k_0$ is the
size of the smallest county in the dataset. The interesting thing to note is
the broadness of the distribution: this type of distribution is often called
``fat-tailed''. \Fref{fig:county1}(c) illustrates the same feature by
log-binning the raw data. The resulting distribution $P(k)$ is also
``fat-tailed'' and, since $C(k)$ is related to $P(k)$ by $\sum_k P(k)$,
\fref{fig:county1}(b) and \fref{fig:county1}(c) basically carry the same
information. $P(k)$ is called the frequency distribution and is normalized
such that $\sum_{k_{0}}P(k)=1$.

\begin{figure}
\begin{center}
\includegraphics[width=0.30\textwidth]{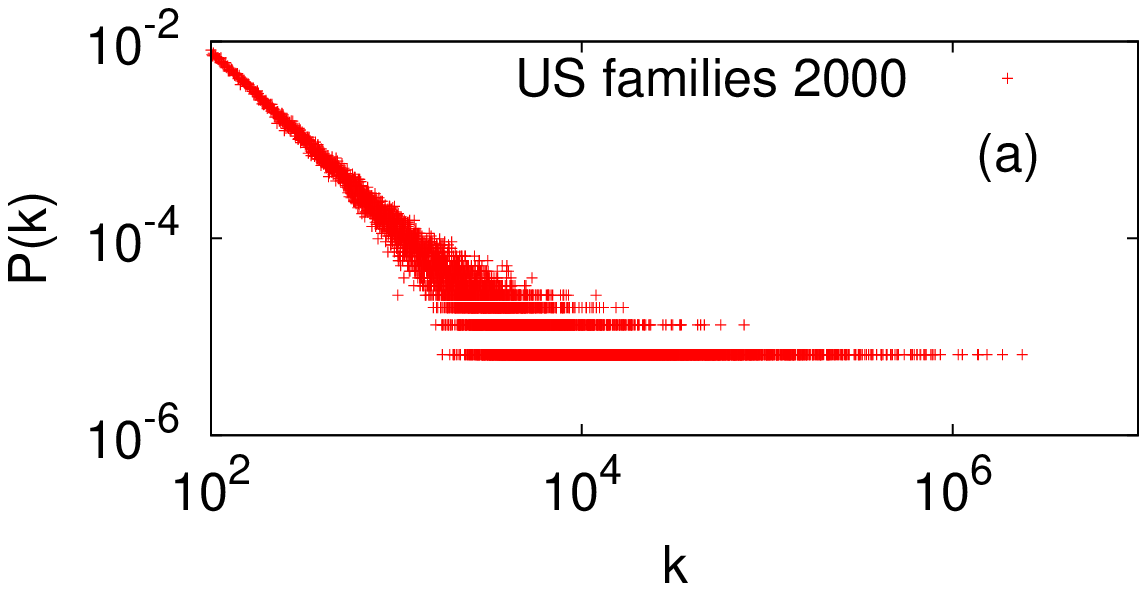}
\includegraphics[width=0.30\textwidth]{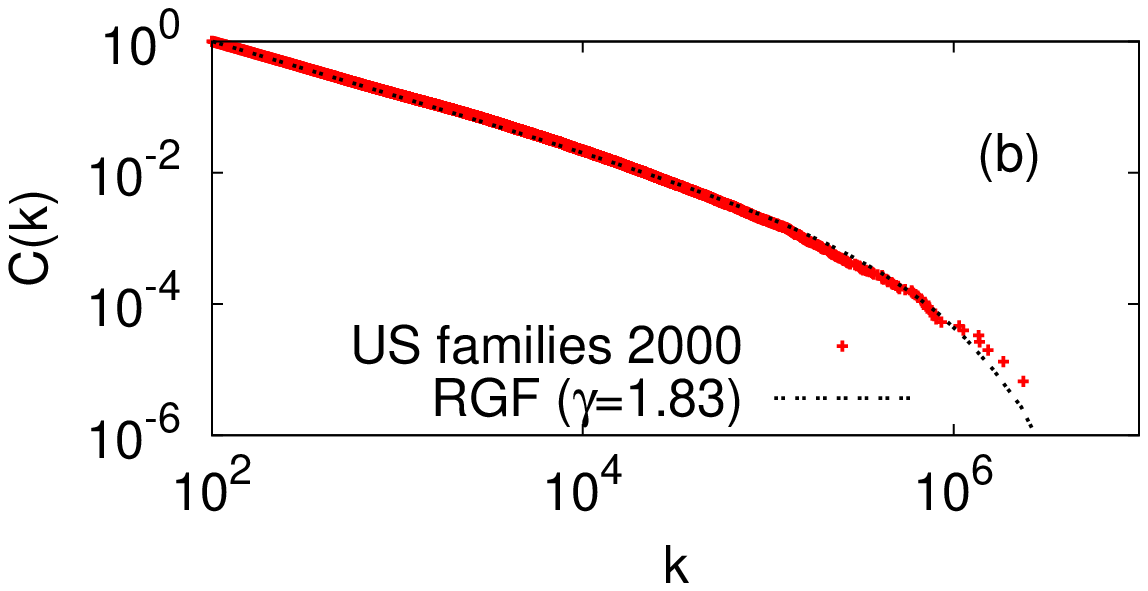}
\includegraphics[width=0.30\textwidth]{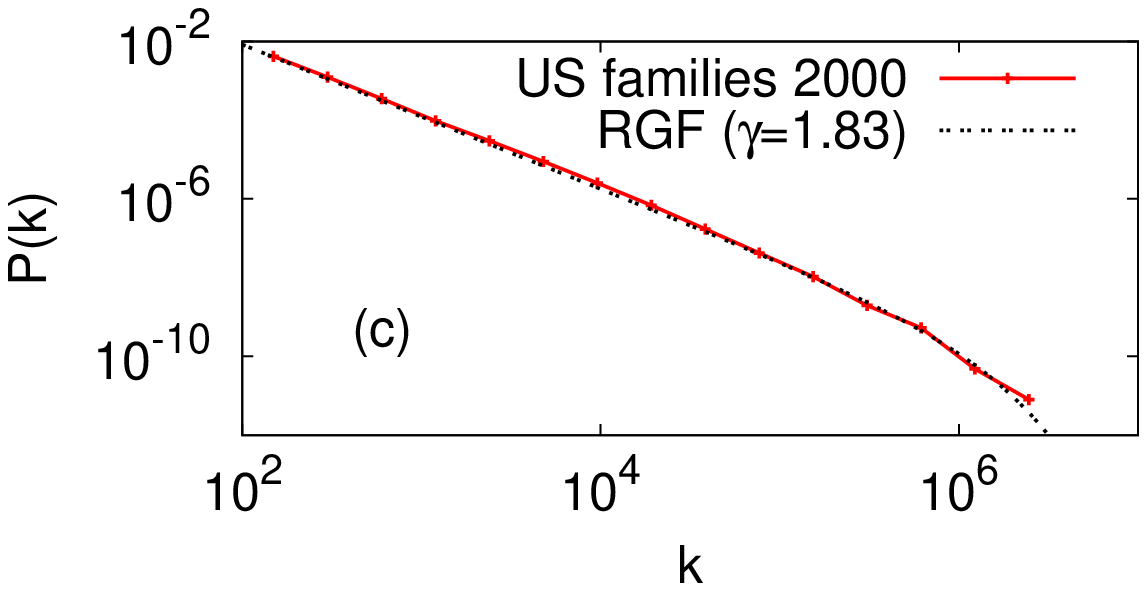}
\end{center}
\caption{  Distribution of US family names
(source: US Census 2000~\cite{usfamily}).
(a) Raw-data in a log-log plot; (b) Cumulative C(k); (c) Binned P(k). Dashed
curves in (b) and (c) give the RGF prediction for the dataset.
}
\label{fig:usfamily1}
\end{figure}

\begin{figure}
\begin{center}
\includegraphics[width=0.30\textwidth]{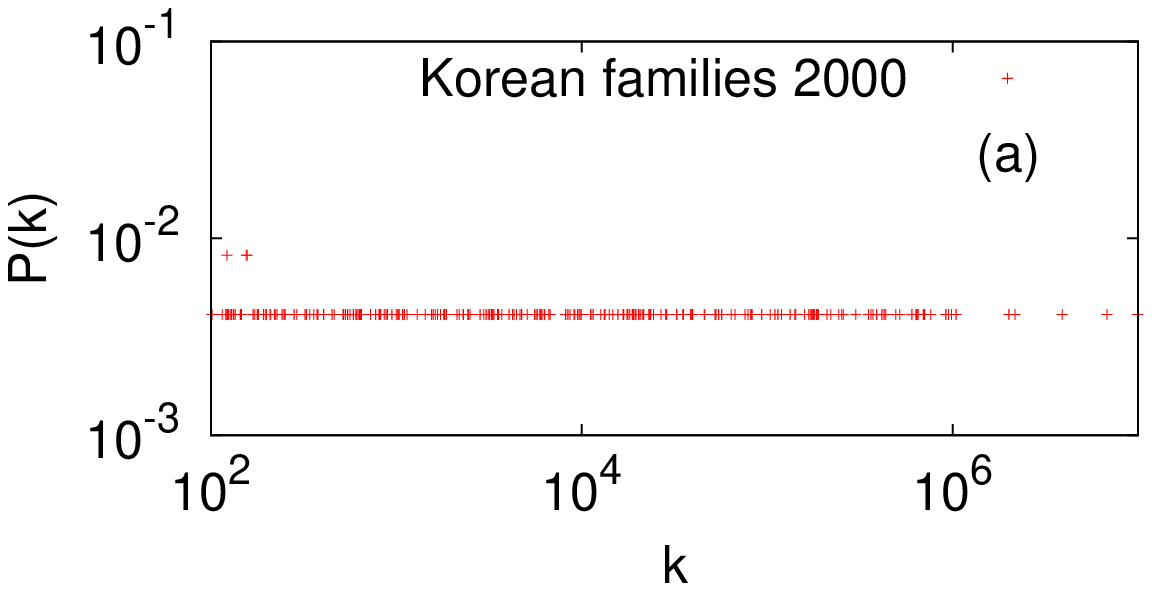}
\includegraphics[width=0.30\textwidth]{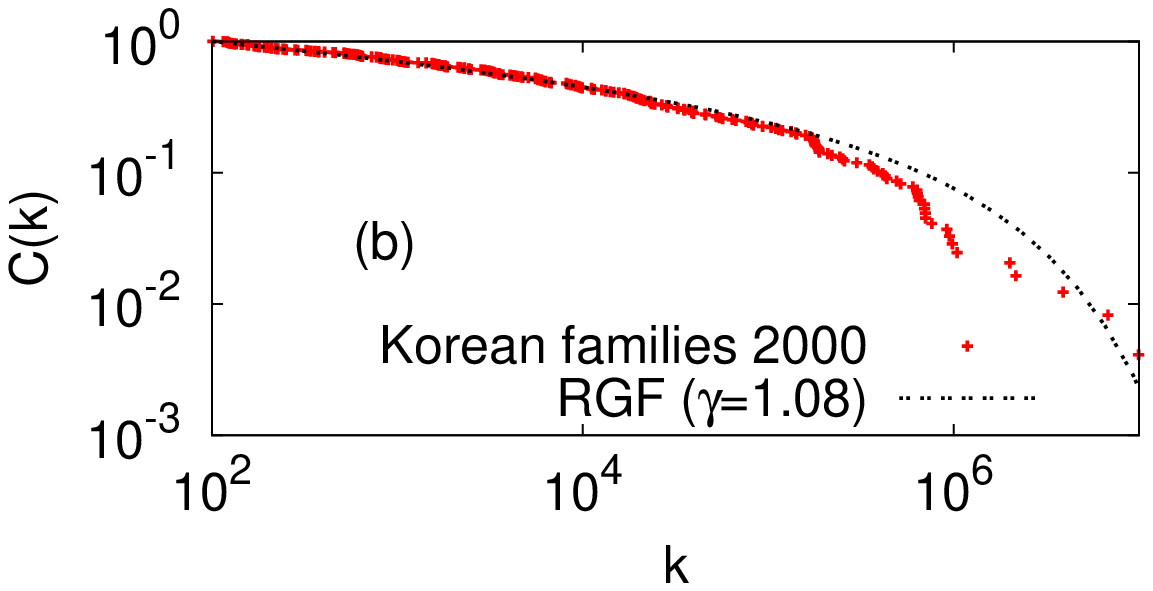}
\includegraphics[width=0.30\textwidth]{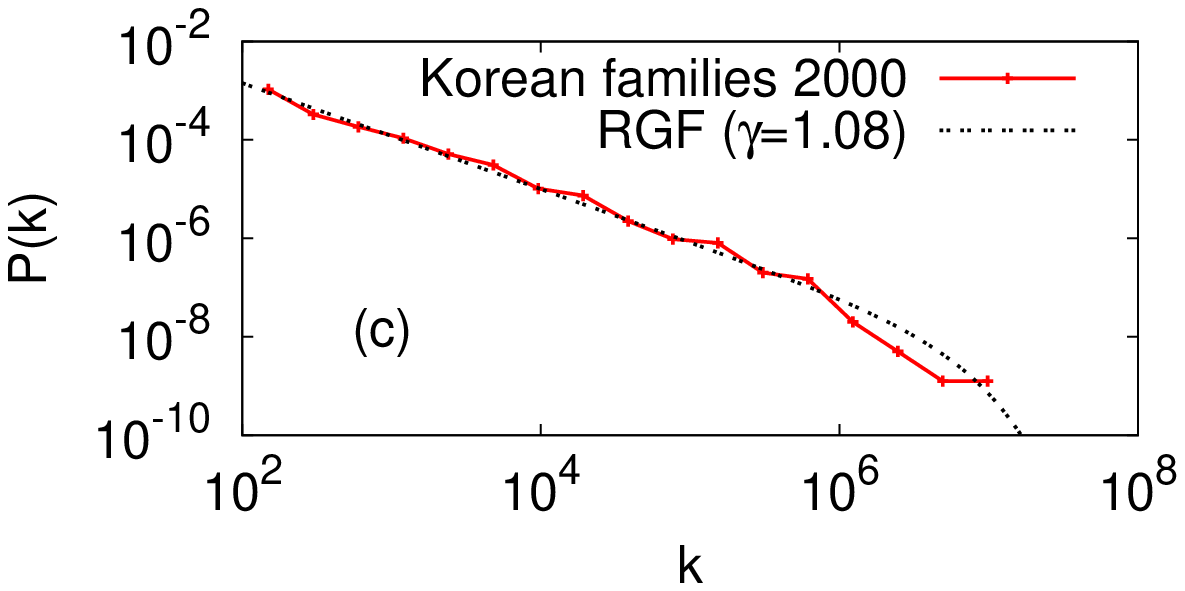}
\end{center}
\caption{Distribution of Korean family names
(source: 2000 South Korea Census~\cite{krfamily}).
(a) Raw-data in a log-log plot; (b) Cumulative C(k); (c) Binned P(k). Dashed
curves in (b) and (c) give the RGF prediction for the dataset.
}
\label{fig:krfamily1}
\end{figure}

\begin{figure}
\begin{center}
\includegraphics[width=0.30\textwidth]{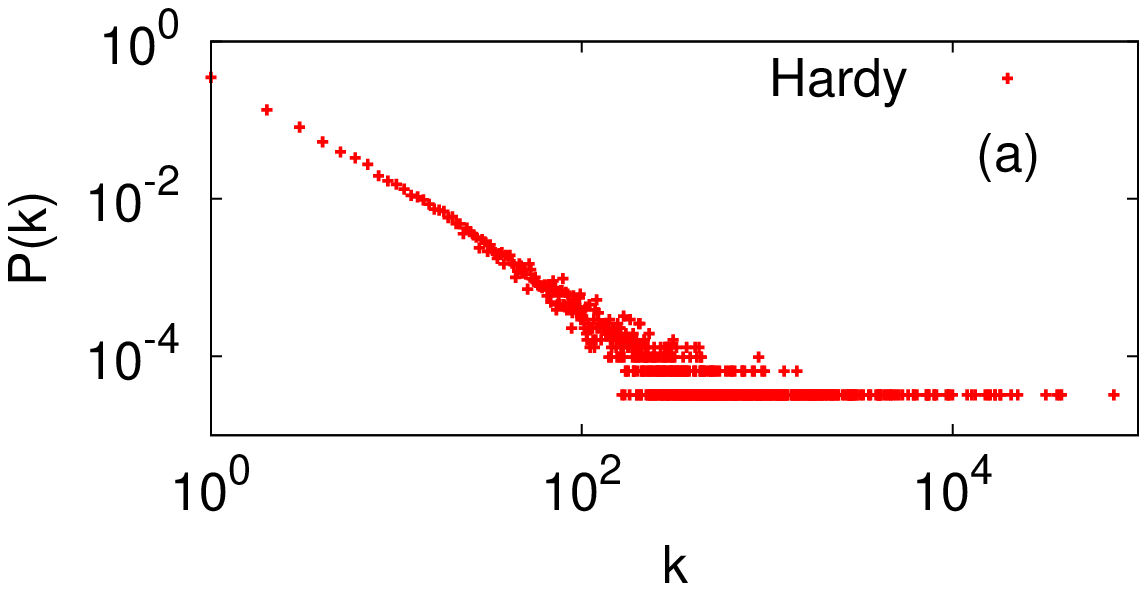}
\includegraphics[width=0.30\textwidth]{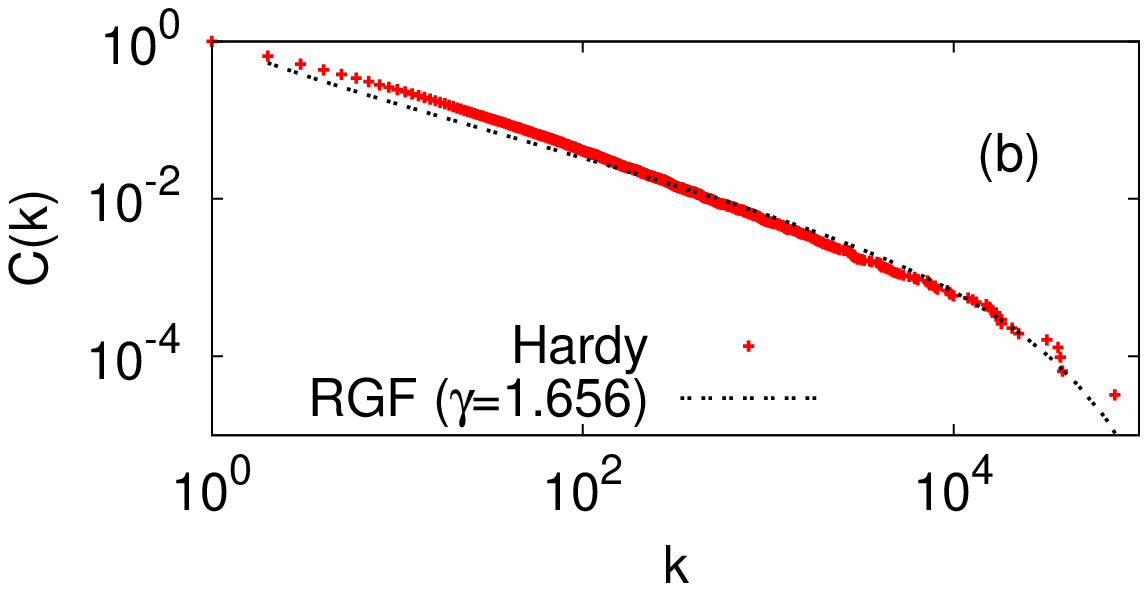}
\includegraphics[width=0.30\textwidth]{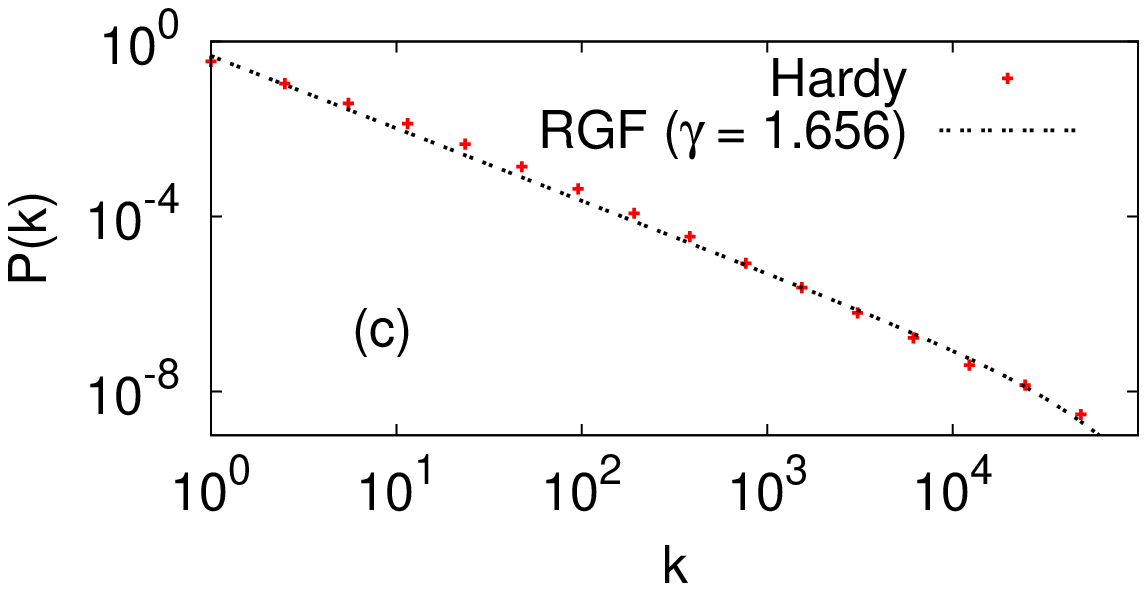}
\end{center}
\caption{Distribution of word-frequencies for the author Thomas Hardy
(source: see Table I of Ref.~\cite{seb09}) (a) Raw-data in a log-log plot; (b)
Cumulative C(k); (c) Binned P(k). Dashed curves in (b) and (c) give the
RGF prediction for the dataset.}
\label{fig:hardy1}
\end{figure}

\begin{figure}
\begin{center}
\includegraphics[width=0.30\textwidth]{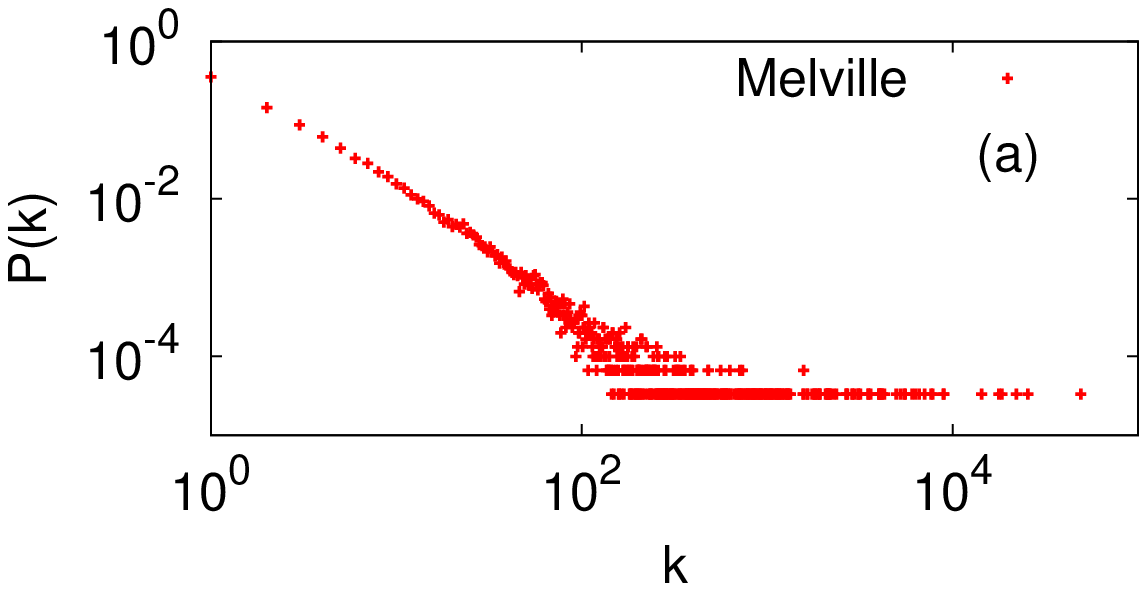}
\includegraphics[width=0.30\textwidth]{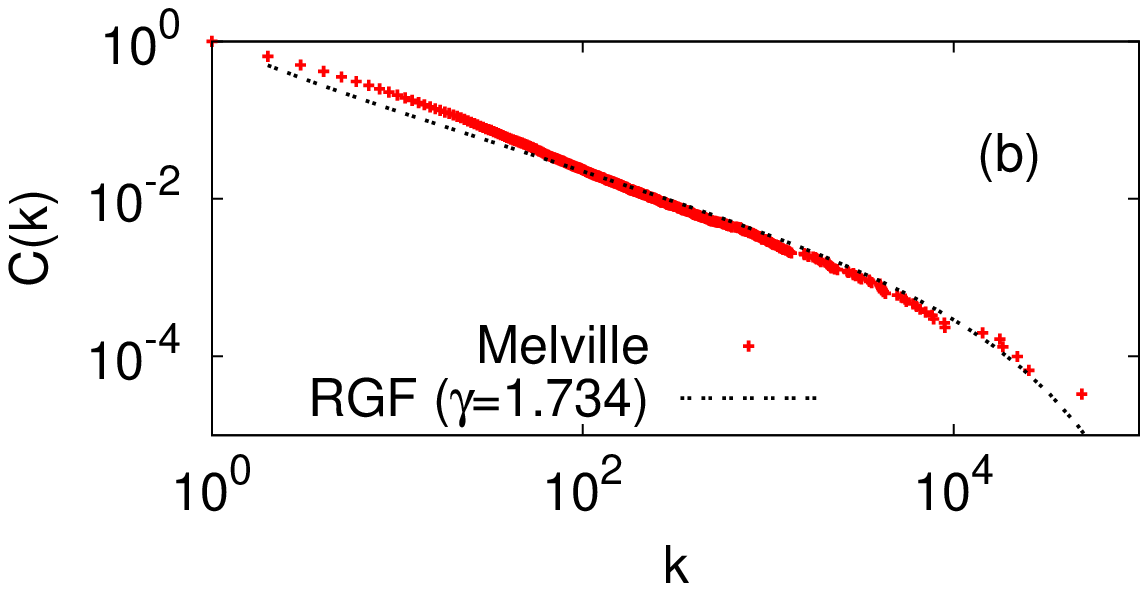}
\includegraphics[width=0.30\textwidth]{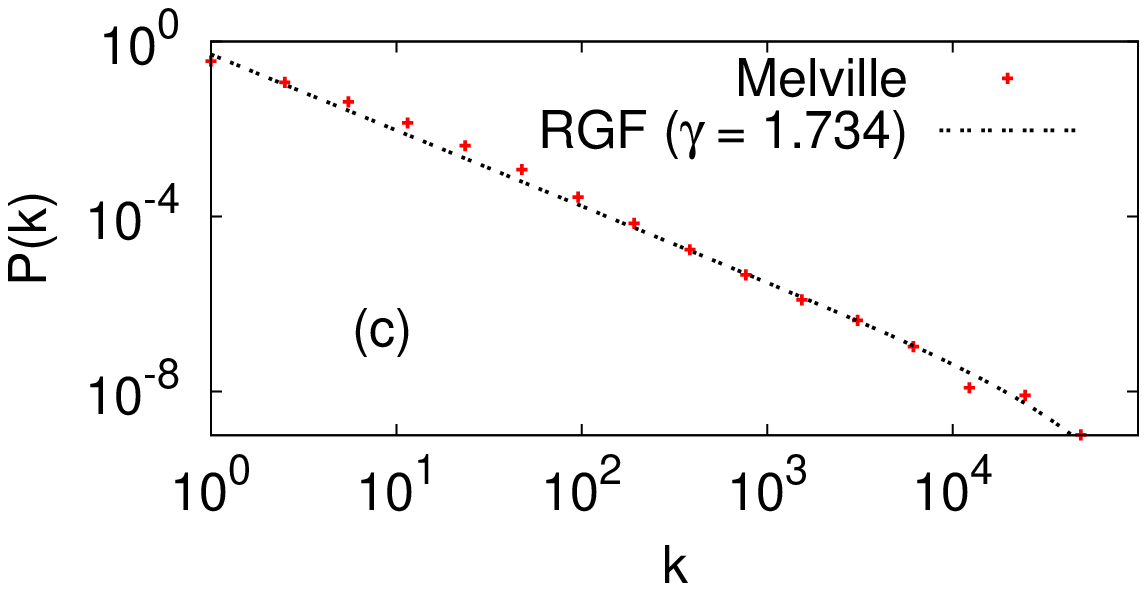}
\end{center}
\caption{Distribution of word-frequencies for the author Herman Melville
(source: see Table I of Ref.~\cite{seb09})
(a) Raw-data in a log-log plot; (b) Cumulative C(k); (c) Binned P(k). Dashed
curves in (b) and (c) give the RGF prediction for the dataset.
}
\label{fig:melville1}
\end{figure}

The first thing one may ask is if this fat-tailed feature is specific to
county sizes in US. The answer that it indeed is a typical feature of
city distributions was first noted by Auerbach in 1913~\cite{auerbach13} and
has since then been amply verified. We illustrate this in
\fref{fig:france1} by including the communal sizes of France~\cite{commune}.
In this case the total
population is $M=5.1 \times 10^7$, the number of communes $N=9011$ and the
largest
commune is Marseille with $k_{\max}=8.5 \times 10^5$ inhabitants (see
\tref{table:data}). The data is displayed in the same way and the similarity
of the shape of the fat-tailed distribution is striking. What is the reason
for this similarity? The first thought might be that it must be connected to
some specific human endeavor of creating towns for reasons related to
fertility, immigration, economics, commerce and defense and hindered by
factors like epidemics, emigration, war, famine and earthquakes. However,
this thought is to some extent superseded by \fref{fig:usfamily1}, which
shows that the family names in US are distributed in a very similar way (data
from US Census 2000~\cite{usfamily}). In this
case the total number of persons in the dataset is $M=2.4 \times 10^8$,
the number of family names is $N=1.5 \times 10^5$, and the most common name
Smith has $k_{\max}=2.4 \times 10^6$ carriers.
Thus ``fat tails'' are not exclusive for city-size
distributions. The second example of family names is from Korea (data taken
from 2000 South Korea Census~\cite{krfamily}). As apparent
from \fref{fig:krfamily1}, this distribution also has a ``fat tail''.
However, the fall-off in the log-log plot is slower than for the US family
names. Nevertheless, a common feature is the fat tails. Perhaps one could
argue that the common factor between city sizes and family names is that in
both cases the basic entity are people and so that the reason could be
linked to some human sociology~\cite{zanette01,baek07}. However,
\fref{fig:hardy1} and \fref{fig:melville1} show that the same fat-tailed
feature remains true for the word-frequency distribution of words of an
author. \Fref{fig:hardy1} shows the data compiled from a large set of
Thomas Hardy's novels (the data set is taken from Table I of
Ref.~\cite{seb09} and is obtained by adding together novels by Hardy into a
single giant novel). In this case, the total number of words is $M=1.3
\times 10^6$,
the number of distinct words is $N=3.0 \times 10^4$ and the most common word
is `the' which appears $k_{\max}=7.4 \times 10^4$ times. Again the same type
of ``fat tailed''
distribution, as for the previous cases, is obtained. The second example of
word-frequency of an authors is Herman Melville (the data is obtained in the
same way as for Hardy and is taken from Table I of Ref.~\cite{seb09}). This
time $M=7.4 \times 10^5$, $N=3.0 \times 10^4$ and the number of `the' is
$k_{\max}=4.9 \times 10^4$. The result is very much the same as for the
other cases.

\begin{figure}
\begin{center}
\includegraphics[width=0.5\textwidth]{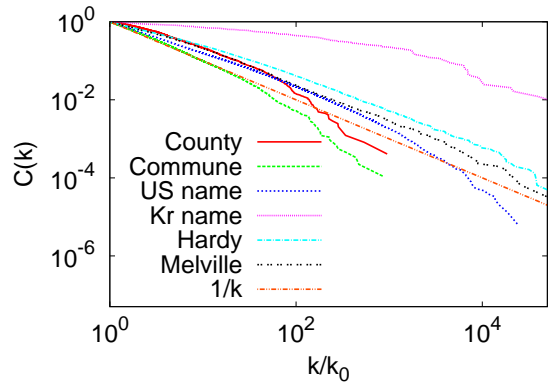}
\end{center}
\caption{Direct comparison of the raw data for the six cases in the
cumulative representation $C(k)$. The horizontal axis is plotted as $k/k_0$
where $k_0$ is the size of the smallest group in respective dataset.
Although the functional forms are clearly different in detail for all the
cases, the ``fat tails'' are a shared feature.}
\label{fig:cum}
\end{figure}

\Fref{fig:cum} gives a direct comparison between the raw data for the
six cases by using the cumulative representation $C(k/k_0)$ where $k_0$ is
the size of the smallest group of the dataset in each case. The
distributions in all cases are "fat tailed". However the precise functional
form differs in every case. Of the pairs for the three different phenomena,
the word frequencies for Hardy and Melville come closest to be similar.
Nevertheless, they are clearly different. The two city-size distributions
are also rather similar but not as close as the two word-frequency
distributions. Finally the two family-name distributions are quite
different. On the other hand, the US family names and the Melville's
word frequency have a substantial overlap for smaller $k/k_0$. The full
drawn straight line in \fref{fig:cum} is the prediction of Zipf's law
$C(k)\propto 1/k$~\cite{Zipf32}. As seen from the figure, Zipf's law does not
give a particular good mathematical description of the data. All the data
sets fall on convex curves in a log-log plot and only the French commune
data follow Zipf's law over a limited interval for smaller $k$. You can
argue that all the data sets to some extent follow a power-law $k^{-\alpha}$
with different power-law indices $\alpha\leq1$ in limited regions for
smaller $k$. In such a case it is only the Zipf value $\alpha=1$ which is
too restricted, so that the more general power-law form could still be a
significant feature. However, the undeniable fact is that all the curves are
somewhat convex and hence that the power-law form does not give a complete
mathematical representation of the data. Nevertheless, one can argue that
Zipf's law catches the essential fact that the distributions are broad and
have fat tails and that the broadness to a first approximation can be
estimated by power-law distributions, albeit with different
power-law indices. However, from such a view point, it is really the
broadness which is the essential thing: the power-law approximations
\emph{per se} may not have any direct bearing on the understanding.
    
There are two possible hypotheses you can start from: either one argues
that the ``fat-tailed'' distributions are essentially system-specific and that
the similarity of the distributions is just accidental and hence of no
particular significance, or you can side with Herbert Simon in
Ref.~\cite{simon55} and argue that the similarities imply an underlying
system-independent stochastic process which accounts for the ``fat tails''.
In this paper, we pursue the latter possibility.
 
\section{Random Group Formation}
\label{sec:rgf}
A common feature of all the data-sets in \sref{sec:what} is that they on an
abstract level can be described as objects divided into categories. The
point made here is that, considering the immense variety of systems which
display the same type of ``fat tails'', it seems hard to imagine any
other general shared feature. The question addressed here is what can be
deduced about the size of the categories solely based on this common
feature.
 
The starting point for the RGF model is as follows:
You have $M$ numbered balls and $N$ boxes. The box sizes are $N(k)$, which
means that a box of size $k$ has $k$ distinct slots which a ball can occupy.
There are hence in total $M$ distinct slots and since you have no other
knowledge, you assume that the probability of finding a specific ball at any
address is equal. This is the Bayesian assumption. This means that the
chance of finding a specific ball at a specific address is $P_{\rm dice}=1/M$.
To make it less abstract, one can consider people divided into towns. A town
of size $k$ has $k$ addresses where a person can live. You can imagine that
the persons move around so as to come close to friends and suitable job
opportunities. However, the motivation and initiative differ quite a lot.
But if you do not have any information of these system-specific driving
forces, your best guess is that any person has an equal probability to live
at any address.

Under this Bayesian assumption of equal address probability, what is the
best estimate you can make for the distribution $N(k)$? One may then note
that $P(k)=N(k)/N$ can be viewed as a probability distribution. This means
that, if you know $P(k)$, then a typical expected $N(k)$ is obtained by
randomly drawing $N$ box-sizes from the probability distribution $P(k)$. The
most likely $P(k)$ corresponds to the maximum entropy of
$S[P(k)]=-\sum_kP(k) \ln P(k)$ under the constraints that $M$ and $N$ are
given, together with the constraint given by the condition $P_{\rm dice}=1/M$.
The last constraint can be handled by maximum mutual
information~\cite{cover06} or equivalently by minimum information cost, as
explained in \ref{app:max}. The information cost enters as follows: the
information to localize a ball with no additional knowledge is
$I_{\rm total}=\ln M$ (in nats). The information needed to localize a ball,
if you know that it is contained in a box of size $k$, is $\ln[kN(k)]$.
This means that if you draw a value $k$ from the probability function
$P(k)$, then $\ln [kN(k)]$ is the information it will cost you to localize a
specific ball belonging to this $k$ value: the
information cost is defined as the additional info which on the average is
needed to localize a ball if you know the box size,
$I_{\rm cost}=\sum_kP(k) \ln[kN(k)]$. The best estimate of $P(k)$ is
obtained by minimizing the information cost subject to the $M$ and $N$
constraints, i.e., minimizing $G([P(k)]$
\begin{equation}
G([P(k)])=I_{\rm cost}[P(k)]+c_1\sum_kP(k)+c_2\sum_kkP(k)              
\label{eq:G}
\end{equation}
where $c_1$ and $c_2$ are positive constants. It is interesting to note that
$G([P(k)])$ can be regarded as the total information cost: each additional
constraint means that information is added to the specification of the
system and hence adds to the cost. The variational solution is
\begin{equation}
P(k)=A\frac{\exp(-bk)}{k}
\label{eq:gamma=1}
\end{equation}
where $A$ and $b$ are determined by the conditions $\sum_{k=1}^MP(k)=1$ and
$\sum_{k=1}^MkP(k)=\left<k\right>=M/N$. This distribution is hence the most
likely distribution provided that you randomly place $M$ numbered balls into
$N$ boxes under the condition that the chance of finding a specific ball in
a specific slot is equal. A crucial observation is that equal chance means
no preference and that any preference means additional knowledge. Additional
knowledge means larger \emph{a priori} knowledge which means smaller entropy
for the distribution $P(k)$. Thus any additional \emph{a priori} knowledge
means smaller entropy. This observation makes it possible to go one step
further without losing generality.

In case of the real systems described in the preceding section, one can
think of innumerable processes involved in creating the data. Among these
there are likely to be processes which breaks the no preference condition
and hence lower the entropy of the distribution $P(k)$. If we \emph{a
priori} assume that the entropy is lowered by the amount $\Delta S$, caused
by the combined effect of all such unknown non-preference breaking
processes, then this can simply be incorporated into the variational
estimate as yet another Lagrangian constraint 
 
\begin{equation}
G([P(k)])=I_{cost}([P(k)])+c_1\sum_kP(k)+c_2\sum_kkP(k)+c_3 S[P(k)] 
\label{eq:GS}
\end{equation}
where $c_3$ is an additional Lagrangian multiplier and the solution is
\begin{equation}
P(k)=A\frac{exp(-bk)}{k^\gamma}
\label{eq:gamma}
\end{equation}
where $\gamma=\frac{1}{1-c_3}$ and the multipliers are determined by the three
conditions $\sum_{k=1}^MP(k)=1$, $\sum_{k=1}^MkP(k)=\left<k\right>$ and
$-\sum_{k=1}^MP(k) \ln P(k)=\Delta S$.

To turn this into a predictive estimate, one also needs an estimate of
$\Delta S$. Here the particular functional form of $P(k)$ given by
\eref{eq:gamma} provides a convenient estimate of $\Delta S$; since
$\Delta S$ lowers the entropy of $P(k)$, it always makes the ``fat tail'' less
broad. This means that it also lowers the value $k_{\max}$. The value of the
member of the largest group is well defined for each dataset and can hence
be used as an input parameter. The connection between $k_{\max}$ and $P(k)$
given by \eref{eq:gamma} can be obtained as follows: determine the value
$k_c$ for which $C(k_c)=1/N$ which means that there is on the average
precisely one box in the interval $[k_c,M]$. The average size of this box
is given by $\sum_{k_c}^MkP(k)=\left<k_{\max}\right>$;
$\left<k_{max}\right>$ is the best possible estimate of $k_{\max}$ for a
dataset generated by the probability distribution $P(k)$. This means that
given $M$, $N$, and $k_{\max}$, the RGF model provides you with a unique
prediction for $P(k)$, where $P(k)$ is obtained from a set of
self-consistent equations. More details on the RGF model are given in
\ref{app:self}.

\begin{figure}
\begin{center}
\includegraphics[width=0.50\textwidth]{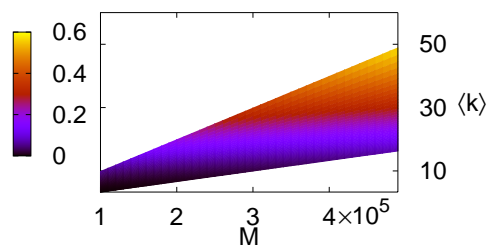}
\end{center}
\caption{The parameters determining the RGF function $\propto
\exp(-bk)/k^\gamma$ for a fixed $\gamma=1.8$: $M$ is given on the horizontal
axis and $\left<k\right>$ on the horizontal right axis, while
$\left<k_{\max}\right>/M$ is color coded. The figure illustrates
that if you know any three of the four parameters $M$, $N$,
$\left<k_{\max}\right>$ and $\gamma$, the fourth is uniquely determined by
the RGF function.
}
\label{fig:contour18}
\end{figure}

\begin{figure}
\begin{center}
\includegraphics[width=0.45\textwidth]{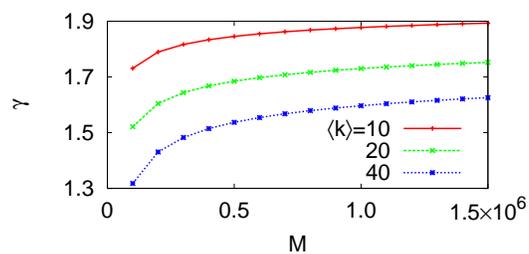}
\end{center}
\caption{Dependence of $\gamma$ on the total number of elements $M$ for
fixed $\left<k\right>=M/N$.}
\label{fig:cross2}
\end{figure}

\begin{figure}
\begin{center}
\includegraphics[width=0.45\textwidth]{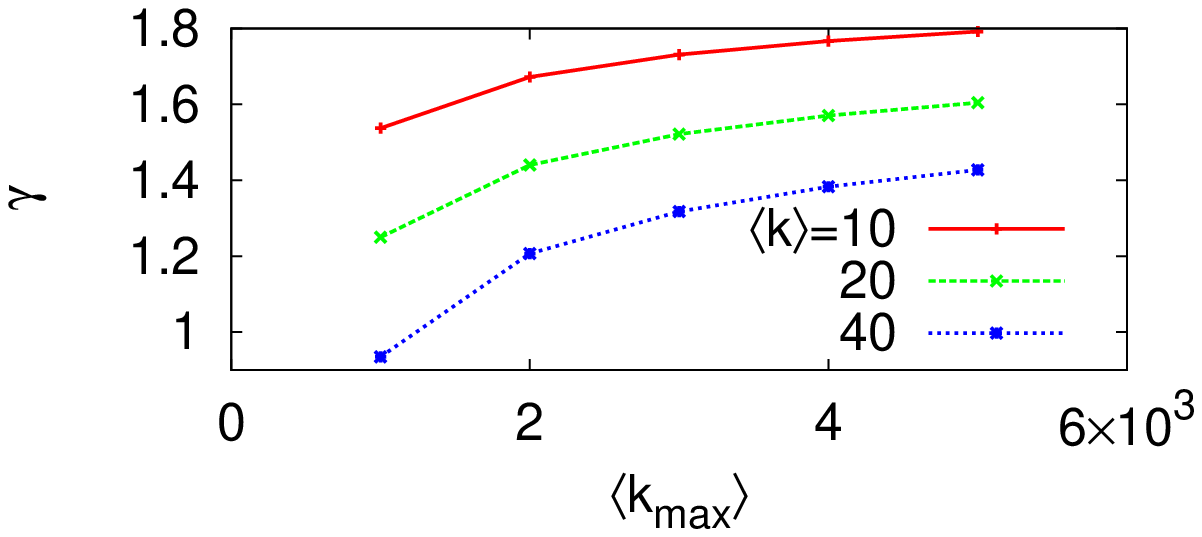}
\end{center}
\caption{Dependence of $\gamma$ on the number of elements in the largest
group, $\left<k_{\max}\right>$, for fixed $\left<k\right>=M/N$.
}
\label{fig:cross3}
\end{figure}

\Fref{fig:contour18} shows the possible solution for a specific value
of $\gamma$ as a function of $M$, $\left<k\right>=M/N$, and $k_{\max}$
(where $k_{\max}$ is color coded). One notes that for any given $M$ and
$\left<k\right>$, there is a whole range of possible $k_{\max}$ values and
this range depends explicitly on the system size $M$.
\Fref{fig:cross2} illustrates how $\gamma$ depends on $M$ for fixed
$\left<k\right>$ whereas \fref{fig:cross3} shows how $\gamma$ depends on
$k_{\max}$ for fixed $\left<k\right>$. \Fref{fig:cross3} is particularly
illuminating because it shows that the power-law index $\gamma$, associated
with the power law which described the distribution for \emph{small} $k$, is
in fact determined by the number of elements in the \emph{largest} group. In
other words, the small $k$ behavior is determined by the non-power-law-like
behavior for large $k$. An interesting consequence of this coupling between
the small and large $k$-dependence is that the power-law index $\gamma$ of
the group distribution increases if one randomly remove a fraction of the
original elements. This will be further discussed in the following section.

As seen in \sref{sec:what}, the data for the group distributions of real
systems often follows a slightly convex function in a log-log-plot. However,
the RGF phenomena in itself does not have this restriction, but it all
depends on the relation between the parameters. Pure power-laws are just
special cases of RGF, which correspond to particular relations between $M$,
$N$ and $k_{\max}$. Also slightly concave distributions are possible within
the general RGF description.

The RGF model leads to a general distribution associated with a minimal
information cost from which the group sizes are drawn. This is somewhat
reminiscent of the Gauss distribution, which is likewise general and
system-independent. One may then ask what the entropy is for a Gaussian (or
Poisson) distribution for a given $M$ and $N$. The answer is that the
Gaussian entropy is smaller than the RGF entropy because the width of a
Gaussian (or Poission) distribution is always smaller or equal to $M/N$,
whereas the reverse is true for the ``fat tailed''. This means that the
Gaussian (or Poisson) form corresponds to a larger information cost; the
process leading to a Gauss curve requires more \emph{a priori} information
to be specified.
From this perspective, the difference in the shapes of the two probability
curves partly stems from the difference in the amount of \emph{a priori}
information needed to specify the global structure of the problem at hand. 

\section{RGF predictions and properties}
\label{sec:predict}
\Tref{table:data} gives the values of $M$, $N$ and $k_{\max}$ for the
raw-data  of the six real examples described in \sref{sec:what}.
This is precisely the raw data needed for uniquely determining the RGF
prediction for each case. In \fref{fig:county1} to \fref{fig:melville1},
these predictions are plotted. The
agreements between the real data and and RGF predictions are remarkably good
in all the cases. It is important to note two things: first, the RGF curves
are predictions based on minimal information. They are \emph{not} any best
fits to the data with some prescribed functional form. It is an important
distinction, because a \emph{prediction} of how the data should be
distributed is conceptually quite different from just fitting a function to
a given data set. The second important thing to note is that the parameter
$\gamma$, which is the counterpart of the power-law index in the cruder
power-law description of the data, is different for different datasets and
is in the range $[1,2]$, where the Korean surname distribution has the
smallest ($\gamma=1.08$) and the French communes the largest
($\gamma=1.98$). \emph{These $\gamma$ values are {\bf not} fitted values,
but are obtained directly from the data in \tref{table:data}.}
If you are an orthodox
believer in power-laws and Zipf's law, you might argue that the data in
\fref{fig:county1} to \fref{fig:melville1} are essentially power laws except
for uninteresting cutoffs at higher $k$ values. Then the thing to note is
that the $\gamma$ values for the RGF curves in 
\fref{fig:county1} to \fref{fig:melville1} are determined by the cutoffs
$k_{\max}$. In other words, the data for French communes, according to the
RGF prediction, falls on a power law with exponent $\approx 2$ for small
$k$ values because the large-$k$ cutoff, Marseille, has about $8.5 \times
10^5$ inhabitants. In order for the French communes to have the same
$\gamma\approx 1.67$ as the counties in US, Marseille would be required to
have about $2.5 \times 10^5$ inhabitants instead. As explained in
\sref{sec:rgf} the cutoff is as essential for the description as is the
number of communes and the total population.

Another consequence of the RGF is that the exponent $\gamma$ for a given
complete dataset with $M$ elements in fact depends on the number of
elements $m<M$ of this dataset which you include in your analysis. To
illustrate this, we choose the word-frequency data from Thomas Hardy:
\fref{fig:hardy1} shows this data, where the full dataset has $M\approx 1.3
\times 10^6$ words and $N\approx 3.0 \times 10^4$ specific words and the
number of the word `the' is $k_{\max}\approx 7.4\times 10^4$ (compare
\tref{table:data}). This information gives
$\gamma\approx 1.66$ and, as seen in \fref{fig:hardy1}(b) and
\fref{fig:hardy1}(c), the RGF prediction gives a very good representation of
the data. Next we randomly remove 99\% of the words so that the total number
of words is instead $m\approx1.3\times 10^4$. The simplest method is just to
randomly remove the words using a computer. It corresponds to a well-defined
mathematical transformation, which in the present context can be called the
Random Book Transformation (RBT)~\cite{seb09,seb10,baayen01}: let the
word distributions before and after the transformation, $P_M(k)$ and
$P_{m}(k)$, be expressed as two column matrices with $N$ elements numerated
by $k$, then
\begin{equation}
\boldsymbol{P}_{m}(k)=C\sum_{k'=k}^{N}\boldsymbol{A}_{kk'}\boldsymbol{P}_M(k')
\label{1}
\end{equation}
where $A_{kk'}$ is a triangular matrix with elements
\begin{equation}
A_{kk'} = \left( \frac{M}{m}-1 \right)^{k'-k}\frac{1}{\left( \frac{M}{m}
\right)^{k'}} \left(\begin{array}{c} k^{\prime}\\k \end{array}\right)
\label{2}
\end{equation}
and $\left( \begin{array}{c}k^{\prime}\\k\end{array}\right )$ is the
binomial coefficient. The coefficient $C$ is
\begin{equation}
C=\frac{1}{1-\sum_{k^{\prime}=1}(\frac{M-m}{M})^{k^{\prime}}P_M(k^{\prime})}.
\label{3}
\end{equation}
\begin{figure}
\begin{center}
\includegraphics[width=0.4\textwidth]{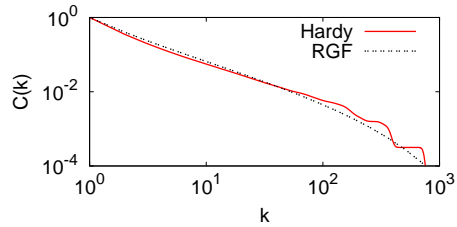}
\end{center}
\caption{Size transformation of the writing by Thomas Hardy. The original
data contains $M\approx 1.3 \times 10^6$ words and is shown in
\fref{fig:hardy1}. The reduced data-set contains 1\% of the words and are
plotted as the full-drawn line in the cumulative plot $C(k)$. This line is an
average over many random removals of 99\% of the words. The RGF prediction is
given by the dashed line. The size transformation causes the power-law index
$\gamma$ to change from $1.66$ (see \fref{fig:hardy1}) to $1.97$ when
only 1\% of the words remain.}
\label{fig:trans}
\end{figure}
More details on the RBT are given in \ref{app:rbt}.
The transformed Hardy has $m=1.3\times 10^4$, $n=3170$ and the number of
`the' is $k_{\max}=742$. Note that the transformations of $M$ and $k_{\max}$
are trivial: both are reduced by a factor of hundred. However, the
transformation of $N$ is nontrivial: the chance of removing a specific word
which occurs $k$ times in the original dataset depends in a nontrivial way
on $k$. The three values $(m,n,k_{\max})$ for the transformed book give a
corresponding RGF prediction for the distribution. This prediction gives
$\gamma=1.9655$. Thus the prediction is that $\gamma$, because of the size
transformation, increases from $1.66$ to $1.97$. This is confirmed by the
actual data for the smaller dataset since the RGF prediction again gives a
very good representation of the data. However, there are some small
deviations. These small deviations are also reflected in a small difference
of the entropy for the transformed data and the RGF prediction: The reduced
data given by the full-drawn curve in \fref{fig:trans} corresponds to
$S=1.49064$, while the RGF prediction corresponds to $S=1.64776$. This means
that the process of randomly removing words imposes some further tiny
constraint in addition to what is absorbed into the RGF prediction. One
should note that, from a system-specific perspective, these small deviations
from the RGF are really the interesting thing, because they do reflect
something system-specific. In the present case, the additional constraint is
a consequence of randomly removing data. However, the most striking thing is
how well the RGF describes the transformation: the removal 99\% of the words
is really a substantial reduction.

\begin{figure}
\begin{center}
\includegraphics[width=0.40\textwidth]{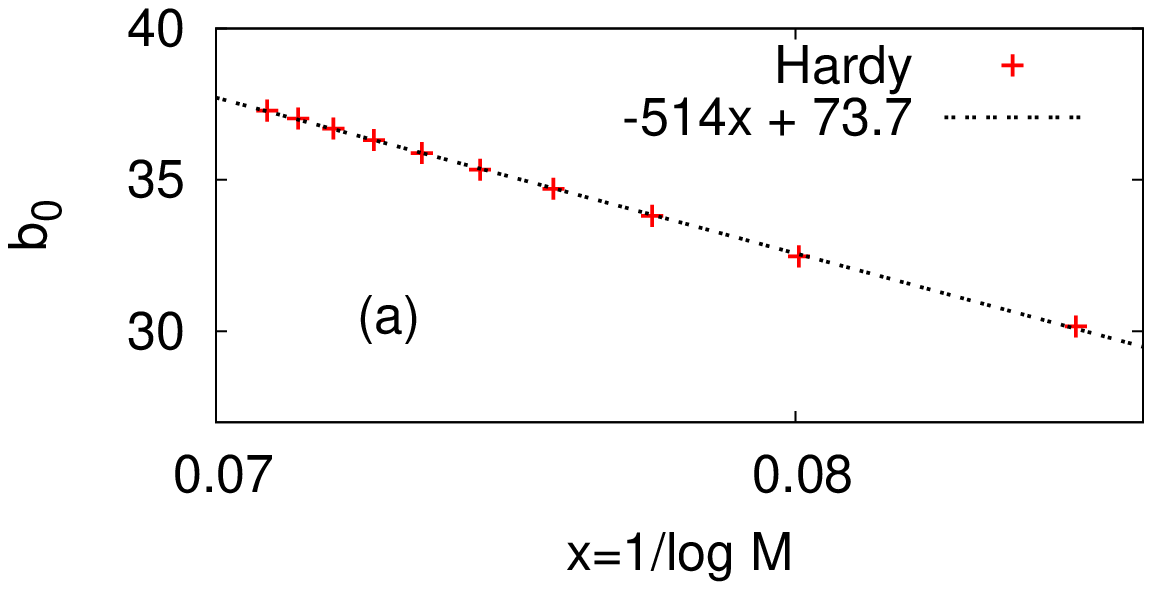}
\includegraphics[width=0.40\textwidth]{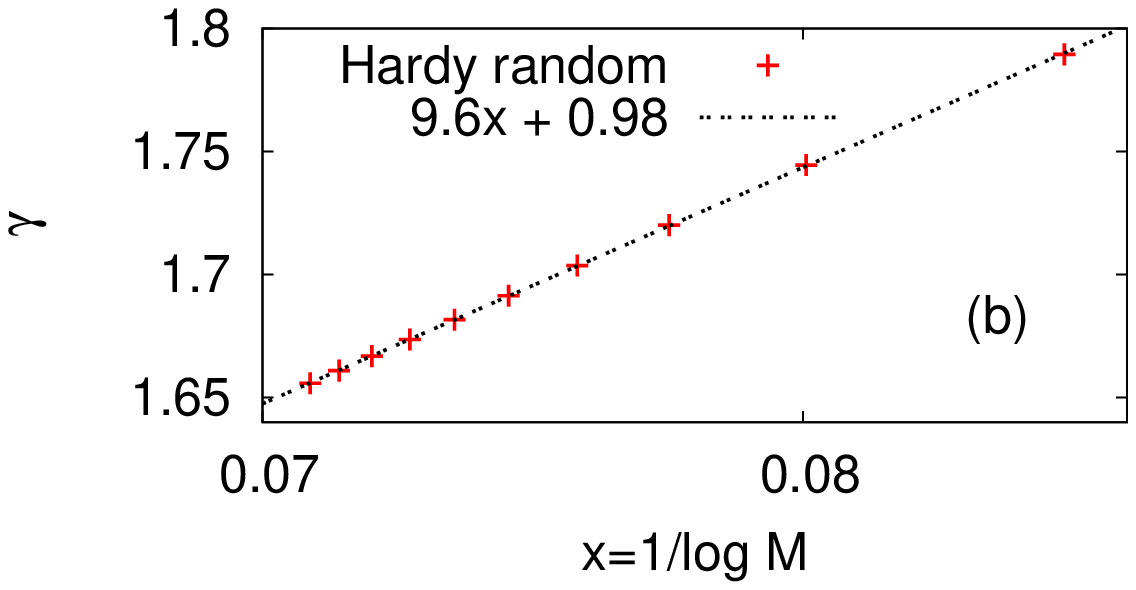}
\end{center}
\caption{Systematic text-length dependence for the writings by Thomas Hardy.
The data is the same as in \fref{fig:hardy1}. The average $\left<k\right>$ as a
function of text length $M$ is obtained by randomly transforming the data
down to a given length. Since the transformation of $k_{\max}$ is trivial,
the knowledge
of $\left<k\right>$ suffices for obtaining the RGF prediction. (a) and (b)
show that the RGF parameters $b_0$ and $\gamma$ are both to good
approximation linear
functions of $1/\ln M$. This makes it possible to extrapolate to $M\rightarrow
\infty$. The extrapolated value of $\gamma$ in the limit $M\rightarrow
\infty$ is in this case $\gamma\approx 1$.}
\label{fig:hardy2}
\end{figure}

\begin{figure}
\begin{center}
\includegraphics[width=0.40\textwidth]{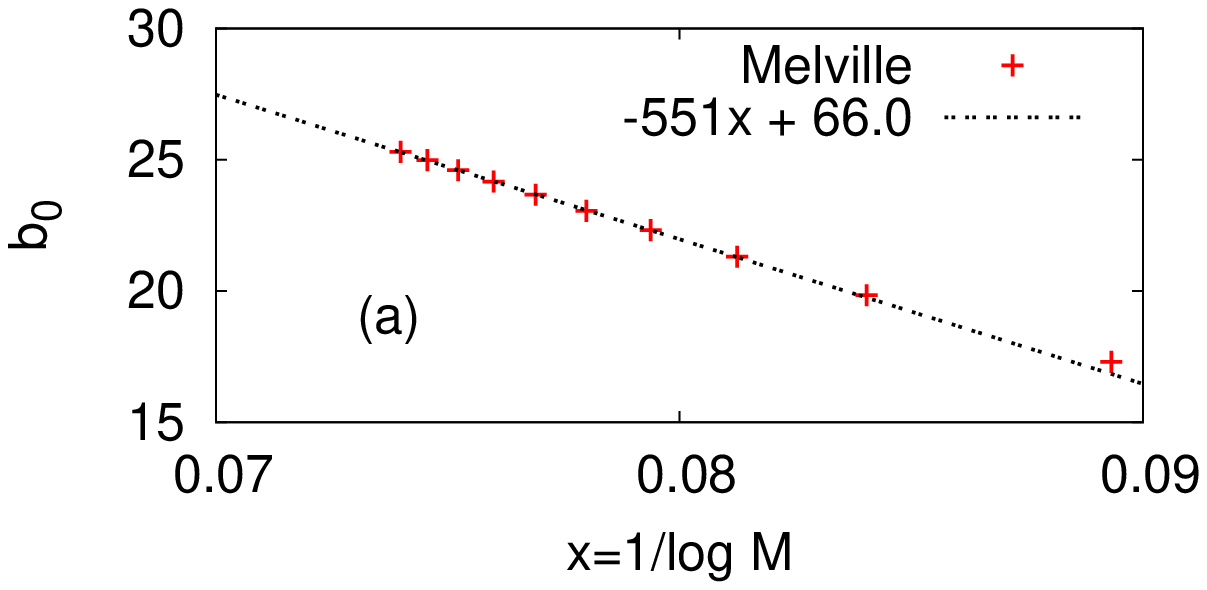}
\includegraphics[width=0.40\textwidth]{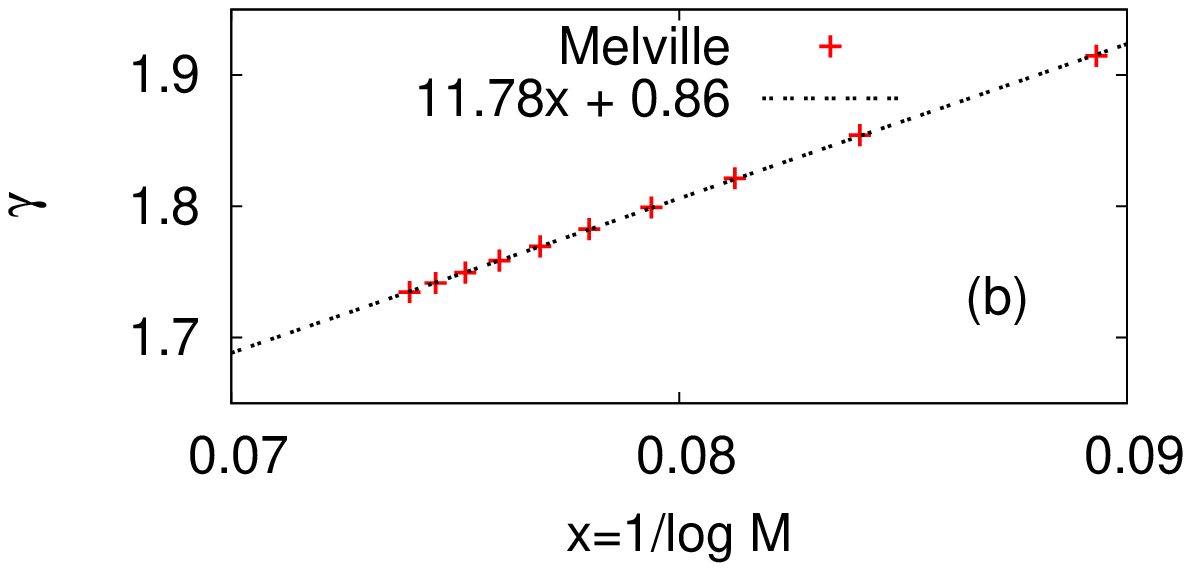}
\end{center}
\caption{Systematic text-length dependence for the writings by Herman
Melville. The data is the same as in \fref{fig:melville1}. The average
$\left<k\right>$ as a function of text length $M$ is obtained by randomly
transforming the data down to a given length. Since the transformation of
$k_{\max}$ is trivial, the knowledge of $\left<k\right>$ suffices for
obtaining the RGF
prediction. (a) and (b) show that the RGF parameters $b_0$ and $\gamma$ are
both to good approximation linear functions of $1/\ln M$. This makes it
possible to extrapolate to $M\rightarrow \infty$. The extrapolated value of
$\gamma$ is $\gamma\approx 0.9$. 
}   
\label{fig:melville2}
\end{figure}

The fact that the power-law index $\gamma$ increases, when the total number
of elements is reduced, also means that $\gamma$ decreases when the number
of elements is increased. One may then ask if $\gamma$ acquires some special
value in the limit $M\rightarrow \infty$. The fact that $\gamma$ decreases
means that the effect of preferential processes diminishes, and from this
perspective one might suspect that the limit value is the non-preferential
value $\gamma=1$. We have tried to estimate this limit in case of the
word-frequency data by Hardy and Melville: starting from the data in
\fref{fig:hardy1} and \fref{fig:melville1}, we first transform the data to
smaller sizes, by randomly removing words, and obtain the $\gamma$ and $b$
for the corresponding RGF. These are plotted as $\gamma$ versus $1/\ln M$ and
$b/M$ versus $1/\ln M$ in \fref{fig:hardy2} and \fref{fig:melville2} for
the data from Hardy and Melville, respectively. The reason why $b_0\equiv
b/M$ is a natural variable is explained in \ref{app:rbt}. As seen in
\fref{fig:hardy2} and \fref{fig:melville2} for Hardy and Melville, the two
quantities $\gamma$ and $b_0$ scale linearly with $1/\ln M$ to a very good
approximation. From this, the limit value of $\gamma$ can be directly
estimated: for Hardy, the value $\gamma\approx 1$ is obtained in the limit
$M\rightarrow \infty$ and for Melville $\gamma\approx 0.9$. This shows that
$\gamma$ does indeed decrease in a systematic way with increasing size and,
furthermore, that the limit value comes close to the non-preferential value
$\gamma=1$. 

We note in passing that in case of a novel written by an author, one may
ask how $\gamma$ changes if one analyzes a small part of the text within
the novel. As shown in Ref.~\cite{seb09}, the result is very similar to the
random removal of words because, to very good approximation, the chance that
a picked word belongs to the frequency class $k$ is on the average
independent of the position in the book.

\section{Relation to Growth Models}
\label{sec:growth} 

Most earlier attempts to explain the broad distributions for
word frequencies, towns and family names are based on growth
models~\cite{simon55,gibrat30,gibrat31,gabaix99,zanette01}. The basic goal of
these attempts focuses on explaining why the data follows a power law with an
exponent close to 2. As seen from the data in \sref{sec:what}, such a power law
does rarely give a good description of the data: the data only approximately
follows a power law for small $k$ and the power-law exponent is usually
significantly smaller than 2. Furthermore, these attempts completely miss
the fact that, as shown here, the number of members of the largest group
determines the power-law index of the power law which approximately
describes the data for small $k$. However, the growth models are also
problematic for two conceptual reasons. The first is that a real growth
model is history-dependent. This is problematic because history and memory
are usually system-specific features and any description which contains such
features is less ubiquitous. The second is the relation between growth,
steady state, and maximum entropy which makes the definition of a growth
model rather flexible.

In order to make the connection between the RGF and growth models, we first
construct a dynamical model which directly leads to the maximum-entropy
solution of the equal-address-probability RGF given by \eref{eq:G}. A
simple dynamical model which achieves this is the following: start with $N$
boxes and $M$ balls and the condition that all boxes must always contain at
least one ball. Then, at each time step, you pick two balls randomly with
equal probability and move one of the balls to the same box as the other.
Any move which attempts to empty a box is abandoned. This dynamical update
has the $P(k)\propto \exp(-bk)/k$ distribution given by \eref{eq:G} as its
steady state solution~\cite{minnhagen07}. Next imagine that you watch this
dynamical process from the vantage point of a single box. This box will then
have a fluctuating number, $k$, of balls between 1 and $M$ following a
trajectory in time $k(t)$. Since the maximum entropy dynamics is completely
ergodic, it follows that $\frac{1}{t_{\max }-t_{1}} \sum_{t_{1}}^{t_{\max
}}k(t_{i})$ $=\left<k\right>$ for $t_{\max} \rightarrow \infty$ and
furthermore that $\frac{1}{t_{\max }-t_{1}}\sum_{t_{1}}^{t_{\max }}\delta
_{k(t_{i}),k}\sim P(k)$ for $t_{\max} \rightarrow \infty$.

\begin{figure}
\begin{center}
\includegraphics[width=0.4\textwidth]{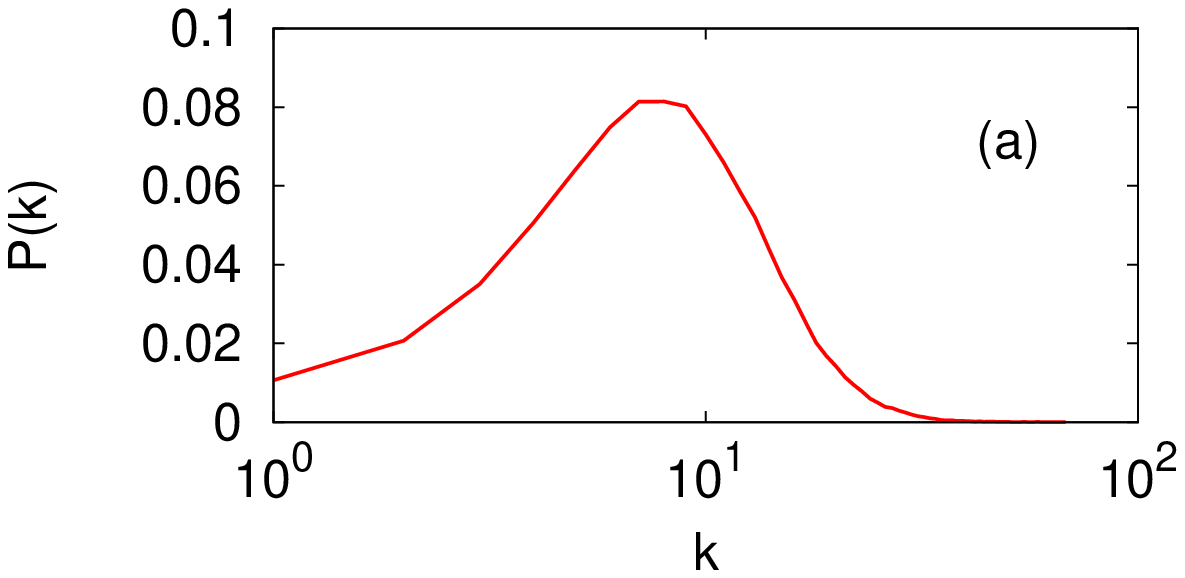}
\includegraphics[width=0.4\textwidth]{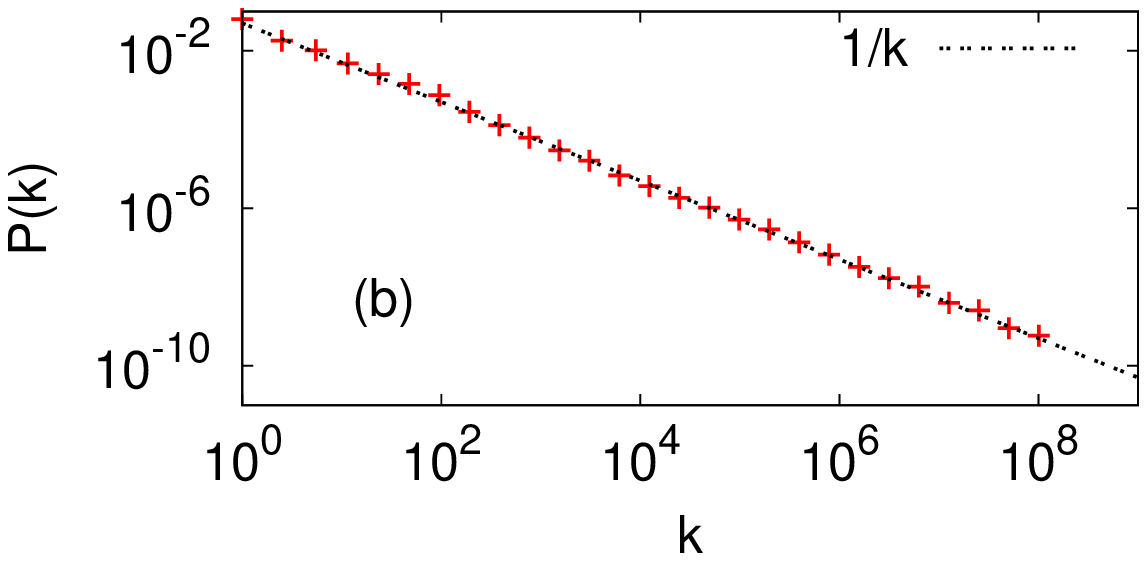}
\includegraphics[width=0.4\textwidth]{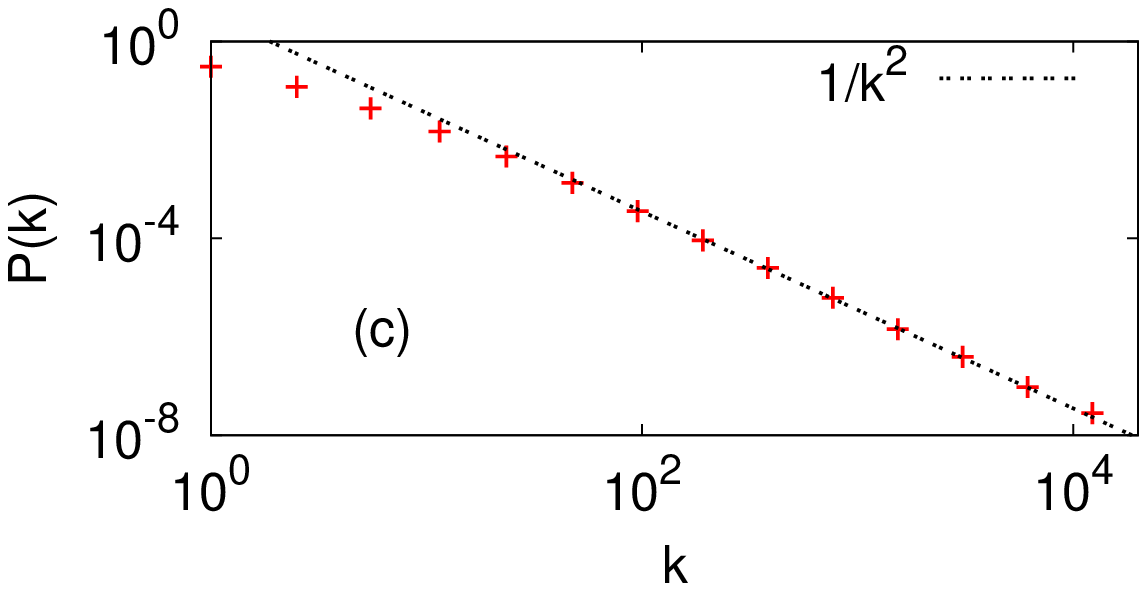}
\includegraphics[width=0.4\textwidth]{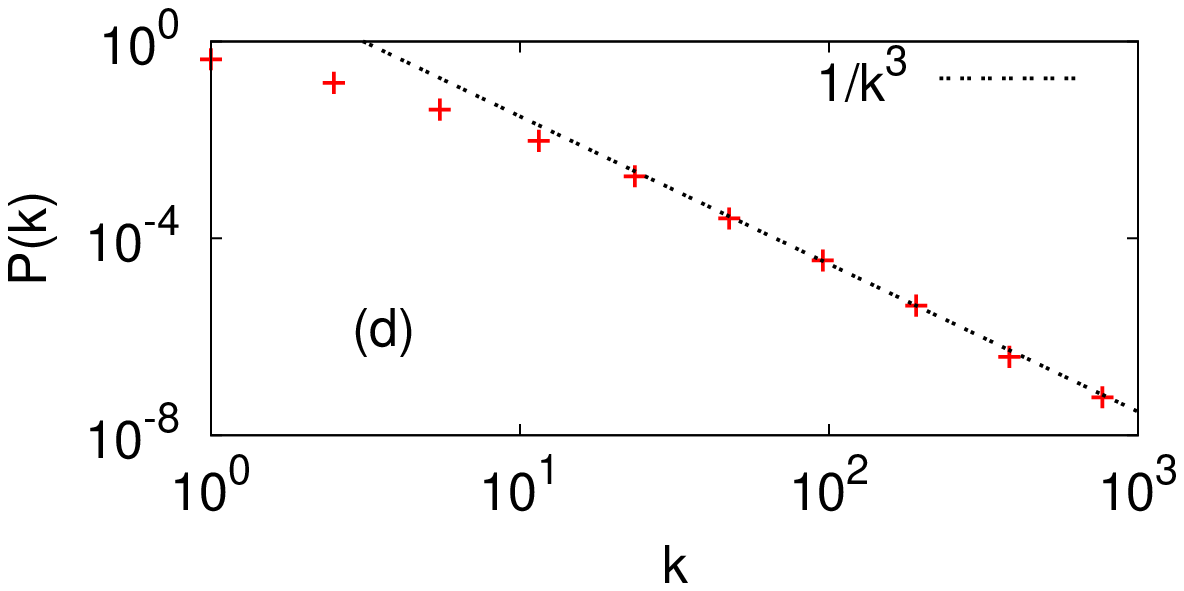}
\includegraphics[width=0.4\textwidth]{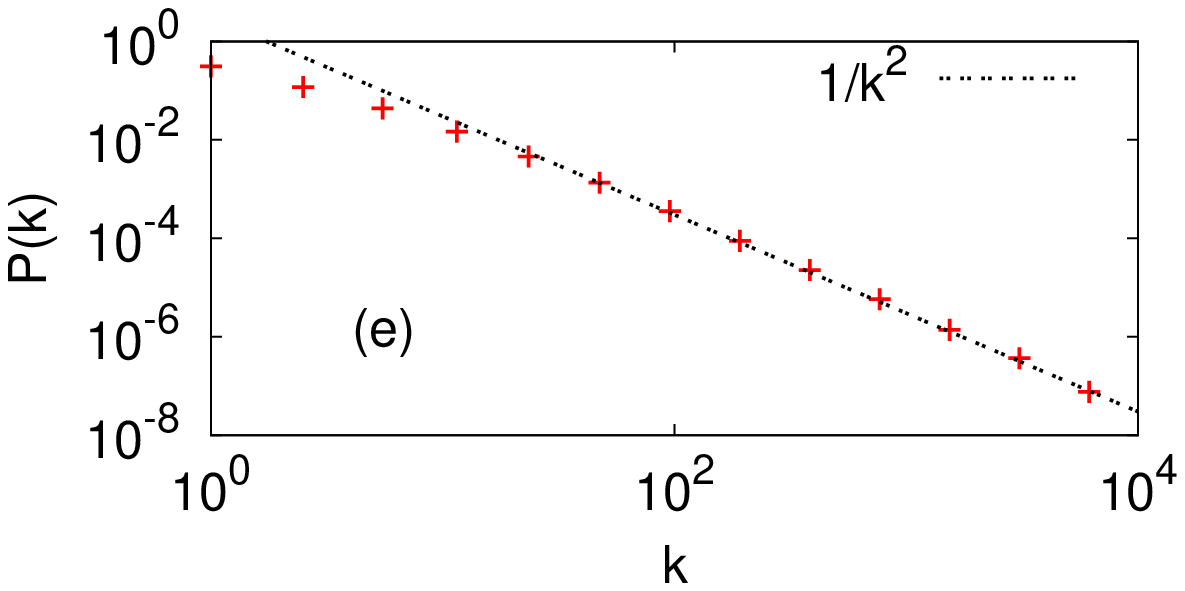}
\end{center}
\caption{(a) Numerical simulation of the discretized version of the Gibrat
model. The outcome of the growth is measure at a fixed time $t$ after start
and the average of many such outcomes gives $P(k)$. The distribution $P(k)$
is close to log-normal. The parameters chosen are $\beta=10k/9$ and
$\alpha=9k/10$. (b) The same model with the same parameters but with the
restriction that the number of balls in the box cannot exceed $M$. This turns
the model into an ergodic model with the steady-state distribution given by
$P(k)\propto 1/k$. (c) Every $k$ can change by $\pm0.1k$. At this point the
model ceases to be a growth model and $P(k)\propto 1/k^2$ (d) Every $k$ can
change by $+0.089k$ or $-0.1k$. This is an ergodic non-growing model and
$P(k)\propto 1/k^{\gamma}$ with $\gamma\approx 3$ (e) The same parameter as
in (a) and (b), but every time one subtracts a number of balls corresponding
to the average increase, i.e., $\frac{1}{2}(\alpha-\beta)k$. In this way, the
stochastic part of the model is turned into the non-growing Gibrat model
given in (c) with the same distribution $P(k)\propto 1/k^2$.}
\label{fig:growth}
\end{figure}

Does there exist a corresponding \emph{single box} stochastic process which
yields the same $P(k)$ as the maximum-entropy dynamical process described
above? A particular class of stochastic models are the stochastic
\emph{growth} models, where the box on the average grows proportional to the
size of the box. The generic type can be described as follows: start with
one ball in the box. Then at each time step, with probability 1/2 you either
increase the box by adding $(\alpha -1)k$ balls or you subtract $(1-\beta
)k$ balls, where $\alpha \geq 1$ and $\beta \leq 1$. The precise meaning
is that you pick the balls in the box consecutively and with chance $(\alpha
-1)$ you add an additional ball and similarly for the subtraction. The
boundary condition is that the box has at least one ball. Thus if $(1-\beta
)k$ is too large to be compatible with the boundary condition, you only
remove all balls but one. At each time step, the box increases on the average
with $\frac{1}{2}(\alpha -\beta )k$ balls. This is a generic discretized
model for growth proportional to the box size. This model is in the
continuum limit called the Gibrat model and the corresponding $P(k)$ has a
log-normal distribution which is distinct from a power law.
\Fref{fig:growth} shows the average $P(k,t)$ for the discretized model
for the values $\alpha =10/9$ and $\beta =9/10$ and $t=10$. This is clearly
not a power law but is close to a log-normal distribution. Comparing this to
the word-frequency data in \fref{fig:hardy1} and \fref{fig:melville1}
shows that the log-normal distribution produced by the stochastic growth
model does not match the data. Consequently, growth models producing
log-normal distributions are not contenders for an ubiquitous explanation of
the ``fat tailed'' distributions presented in \sref{sec:what}.

The model can be turned into an ergodic version by imposing a maximal size $M$;
any attempt to increase the box beyond this size is abandoned and a new
attempt is made at this time step. This is just like having a fixed number
of balls which you try to put into the box. The balls which are not in the
box are outside on the table and you choose randomly from them when adding
balls to the box. Every time you remove balls from the box you put them
together with the ones on the table. \Fref{fig:growth}(b) shows that the
stochastic steady state is $P(k)\sim 1/k$. This means that a model, which
grows proportional to the box size at the same time obeys the
condition $\alpha \cdot \beta =1$, has a steady-state version which
corresponds to the maximum-entropy solution. This is just saying that also
for the steady-state single-box-growth model, the chance of finding a
specific ball in the box when it has size $k$ is independent of $k$. The
reason for this can be traced to the particular stochastic update which in a
logarithmic scale corresponds to $\ln k-\ln \alpha \leftarrow \ln k
\rightarrow \ln k+\ln \alpha $. Thus the system wanders randomly among the
values $\ln k$ within the interval $[\ln 1,\ln M]$ and, since there is no
preference, the probability to find the system in any of these points are
equal (modulo a slight correction imposed by the boundary points), from
which $P(k)\sim 1/k$ follows. Next we consider the situation when $\alpha
\cdot \beta >1$ but still a growth model, so that $\alpha -\beta >0$. This
means that $1/\alpha <\beta <\alpha$. In this case, the steady-state
solution instead becomes $P(k)\sim 1/k^{\gamma}$ where $1<\gamma <2$. The
limit case $\alpha =\beta $ is no longer a growth model, but is an
equilibrium model with a well-defined non-growing average size
$\left<k\right>$. In this case, the exponent is $\gamma =2$ as illustrated
in \fref{fig:growth}(c).
When $\beta >\alpha $ the generic model is a non-growing steady-state
model with the solution $P(k)\sim 1/k^{\gamma}$ with $\gamma >2$ as
illustrated in \fref{fig:growth}(d). 
The Gibrat model is often connected to the non-growing steady-state solution
$P(k)\sim 1/k^{2}$ by changing it into a non-stochastic growing part on top
of which a stochastic non-growing part is added~\cite{gabaix99}: at each
time-step one subtracts the number of balls which corresponds to the average
increase during one time step, i.e., $\frac{1}{2}(\alpha -\beta )k$. In this
way the growing model is transformed into a steady average growth on top of
which is added the stochastic model with $\alpha'=\beta'=\frac{1}{2}(\alpha
+\beta )k$. This changes the log-normal distribution into the non-growing
steady-state distribution $P(k)\sim 1/k^{2}$, as is illustrated in
\fref{fig:growth}(e). It is interesting to note that the distribution
$P(k)\sim 1/k^{2}$ has little to do with the \emph{growing} proportional to
the size, but is in fact associated with the corresponding equilibrium
non-growing situation. Thus from a conceptual point the difference between
the log-normal and the $1/k^2$ distributions is precisely the difference
between a stochastically \emph{growing} model and a stochastically
non-growing steady-state solution. 

It is also interesting to note that one could equally well transform the
Gibrat model by instead subtracting $ck$ balls at each time step where
$c>\frac{1}{2}(\alpha -\beta )$. This again yields a model consisting of a
steady average growth and a stochastic non-growing part. However now the
distribution becomes $P(k)\sim 1/k^{\gamma }$ with $\gamma >2$. Thus the
steady-state solutions of the stochastic growth model does give rise to
power laws with a wide range of power law indices, but the actual
\emph{growth} is not responsible for this. As illustrated in
\fref{fig:growth}, starting from the Gibrat model
with a log-normal distribution, you can, by manipulating the boundary
conditions, turn it into effective steady-state solutions which are of
power-law forms and can have a broad range of power-law indices. However, a
general principle, of which manipulation connects to which set of real data,
is lacking.

The Gibrat models are a model for size-proportional stochastic growth of an
\emph{independent} box. The Simon model is a model of proportional growth
for interdependent boxes~\cite{simon55}. It is associated with suggestive
descriptions like ``Rich-gets-richer'' models and ``Preferential
attachment'' models~\cite{mitzenmacher03,newman05,barabasi99,newman06}. In the
context of written texts the generic form can be described as
follows~\cite{simon55}: when you write a text, you either choose a new word or
you repeat one of the words you have used earlier in the text. The Simon
assumption is that you with probability $\alpha$ write a new word and with
probability $1-\alpha$ you repeat an old word chosen uniformly among the
words already written. Within the box-and-ball model, each new word defines
a new box and a word is added to an existing box in proportion to its size.
The average size of a box becomes $\left<k\right>=1/\alpha$ and the
distribution for large $M$ is a power law with exponent
$\gamma=1+1/(1-\alpha)>2$~\cite{simon55}. For the texts by Hardy and
Melville presented in \fref{fig:hardy1} and \fref{fig:melville1}, the
Simon model predicts the power-law indices $\gamma=2.02$ and $\gamma=2.04$,
respectively. As seen from the figures, the Simon model fails to describe
the word-frequency data. Only the data for the French communes in
\sref{sec:what} could be argued to be partially described by a power law
with a $\gamma$ close to 2 in the region of small $k$. But since the Simon
model fails for the US county data and all the other datasets in
\sref{sec:what}, our conclusion is that the Simon model does not have the
ubiquitous generality necessary to explain the ``fat tail'' phenomena.

The lack of generality of the Simon model is to some extent reflected in
Ref.~\cite{mandelbrot59} by Mandelbrot's comment that ``\emph{this is a fairly
reasonable assumption in the case of word frequencies, since a text is
indeed generated word by word. But a national income is surely {\bf not}
distributed dollar by dollar}''. However, the Simon model is in fact also
conceptually unreasonable for texts. This is because it is a true growth
model and hence forces a history dependence on the text which is
incompatible with real texts~\cite{seb10}: since new words are added and old
words repeated at each time-step, the consequence is that the words in a
Simon book which occur only a few number of times in the book occurs more
often at the end of the book. In a typical text, about half the words only
occur once, and in the Simon book, these words are with larger probability
found at the end. In a real text, the words of any frequency group are to
good approximation randomly spread through the book: the history dependence
of the Simon model is a too strong system dependent assumption to make it a
contender for ubiquity~\cite{seb09}.

\section{Summary}
\label{sec:summary}
``Fat tails'' are common features of datasets encountered in very different
contexts. The question is then, if there is a different system-specific
explanation in each case, or if the ``fat tails'' represent an ubiquitous
non-system-specific feature. In this paper, we present evidence for a
ubiquitous explanation based on a Random Group Formation (RGF) phenomena. The
RGF phenomena lead to an explicit prediction of the group sizes for given
values of the total number of elements, groups and the number of elements
in the largest group. As a consequence, the power-law index of the power
law, which approximately describes the data for small $k$, is in fact
determined by the size of the largest group. These predictions were tested
against six large datasets for three system types, i.e., population
distributions, surname distributions and word-frequency distributions.
Two datasets for each type was chosen in order to be able to compare inter-
and intra differences between the datasets. In addition, the datasets were
chosen to be very large in order to get good statistics. The RGF prediction
was found to describe the data very well in all the cases. The RGF phenomena
were also found to be consistent with a systematic change in the power-law
index with system size. This system-size dependence was explicitly
demonstrated in case of the word-frequency distributions.

It was also pointed out that alternative attempts to explain the ``fat
tails'' based on growth models, like the Simon model or the Gibrat model,
give power-law indices larger than 2, whereas the data presented typically
have smaller values. In addition, the growth models can neither explain the
coupling between the largest group and the power-law index, nor the fact
that the power-law index changes in a systematic way with the system size.
The growth models typically give size-independent power-law indices. The
problem with system-specific memory effects for growth models, like the
Simon model, was also pointed out. 

The present investigation leads to the conclusion that a ubiquitous
explanation must account for the fact that the largest group determines the
power-law index describing the small $k$ part of the distribution, as well
as the fact that the power-law index in a systematic way depends on the
system size. For example, a short novel written by an author has a different
power-law index than a much longer novel~\cite{seb10}.

This leaves the critical reader with two options: either one could argue
that the agreements found in the present paper are purely accidental and
that there is indeed no ubiquitous explanation of the ``fat tails''. Or you
could argue that there is a ubiquitous explanation but it is not given by
the RGF. In the latter case, one would then have to come up with an
alternative explanation which accounts for the fact that the size of the
largest group determines the power-law index for small $k$ and which, at the
same time, is consistent with a systematic size dependence.

For our part, we think that the evidence in favor of the RGF explanation is
entirely convincing. Furthermore, since the RGF gives explicit predictions,
its validity is open to further tests.

\ack
We are grateful for support from the Swedish Research Council through grant
no 621-2008-4449.

\section*{References}

\appendix
\section{Minimum information cost and maximum entropy}
\label{app:max}

Suppose you have two variables $x$ and $y$ distributed according to the
corresponding probability functions $p(x)$ and $p(y)$, respectively. The
total entropy $H[p(x),p(y)]$ is then given by
\begin{equation}
H[p(x),p(y)]=-\sum_{x}\sum_{y}p(x,y)\ln p(x,y),
\end{equation}
where $p(x,y)$ is the joint probability for the variables $x$ and $y$. If
the two distributions are independent (so that the probability for a value
$x$ is independent of the value $y$), then $p(x,y)$ reduces to
$p(x,y)=p(x)p(y)$ and the entropy reduces to $H[p(x),p(y)]=H[p(x)]+H[p(y)]$
where $ H[p(x)]=-\sum_{x}p(x)\ln p(x)$. In many situations, $p(x)$ and
$p(y)$ are dependent so that $p(x,y)\neq p(x)p(y)$ or equivalently the
constrained probability $p(x|y)$ (the probability for a $x$ for a fixed
given $y$) is in fact not equal to $p(x)$ i.e. $p(x|y)\neq p(x)$ (note the
general relation $p(x,y)=p(x|y)p(y))$. We here consider the special case
when the distribution $p(y)$ is \emph{a priori} known. In such a case, the
maximum entropy $H[p(x)]$ is obtained by minimizing the constrained entropy 
\begin{equation}
H[p(y)|p(x)]=-\sum_{x}\sum_{y}p(x,y)\ln p(y|x)
\end{equation}
This follows from the general maximum-mutual-information
principle~\cite{cover06}. The mutual information is defined by
\begin{equation}
I[p(x),p(y)]=\sum_{x}\sum_{y}p(x,y)\ln \frac{p(x,y)}{p(x)p(y)}
\end{equation}
and the $p(x)$ corresponding to maximum entropy is obtained by maximizing
the mutual information for the given $p(y)$. However, the mutual information
can also be expressed as $H[p(y)|p(x)]=-I[p(x),p(y)]+H[p(y)]$ and since
$p(y)$ is \emph{a priori} known it follows that maximizing $I[p(x),p(y)]$ is
equivalent to minimizing the constrained entropy $H[p(y),p(x)]$. This
constrained entropy we term the information cost
$I_{\rm cost}[p(x)]=H[p(y),p(x)]$. In the case of the equal-address RGF, the
information cost is given by 
\begin{equation}
I_{\rm cost}[P(k)]=H[p_{i}|P(k)]=-\sum_{k}\sum_{i}p(k,i)\ln p(i|k)
\label{cost}
\end{equation}

The \emph{a priori} known distribution is the equal probability for each of
the $M$ addresses, so that $p_{i}=\frac{1}{M}$. This means that
$p(i|k)=\frac{\delta_{k'(i),k}}{kN(k)}$ and consequently $p(i,k)=\frac{\delta
_{k'(i),k}}{kN}$. Inserting this in \eref{cost} gives
\begin{eqnarray}
I_{\rm cost} &=& -\sum_{k}\sum_{i}p(k,i)\ln p(i|k) =-\sum_{k}P(k)\ln
\frac{1}{kN(k)}\nonumber\\ &=&\sum_{k}P(k)\ln kN(k)
\end{eqnarray}
The conceptual advantage with the quantity $I_{\rm cost}$ is that it has a
simple interpretation: if you know that a specific ball can be found among
the boxes which contain $k$ balls, then $\ln kN(k)$ is the information (in
nats) needed to specify at which specific address the ball can be found. The
average information cost needed for localizing a ball with a known $k$-value
is hence $I_{\rm cost}=\sum_{k}P(k)\ln kN(k)$.

This is an information cost in the sense that the additional info needed to
specify the outcome of the $p(i)$ is no longer available for the entropy
associated with $P(k)$.

\section{Self-consistent Equations}
\label{app:self}
The RGF-curve is obtained by minimizing the total information cost $G[P(k)]$
given by \eref{eq:GS}. The minimum condition $\frac{\partial
G([P(k)]}{\partial P(k)}=0$ leads to the condition
\begin{equation}
\ln P(k)+\ln k +1 +c_1 + c_2k -c_3 \ln P(k) - c_3=0 
\end{equation} 
with the solution given in \eref{eq:gamma} 
\begin{equation}
P(k)=A\frac{\exp(-bk)}{k^\gamma}
\end{equation}
with $\gamma=1/(1-c_3)$, $b=c_2/(1-c_3)$ and
$A=\exp\{-(1+[c_1/(1-c_3)]\}$. The
constants $A$, $b$ and $\gamma$ are determined by simultaneously fulfilling
three conditions. The first two of these conditions are
$\sum_{k=k_0}^MP(k)=1$ and $\sum_{k=k_0}^MkP(k)=\left<k\right>=M/N$, i.e.,
\begin{equation}
\left \{ \begin{array}{l}
\sum_{k=k_0}^{M} A\frac{e^{-bk}}{k^\gamma}=1\\
\sum_{k=k_0}^{\infty}A k\frac{e^{-bk}}{k^\gamma}=M/N
\end{array} \right.
\label{eq:basic}
\end{equation}
where $k_0$ is the size of the smallest box. The natural limit is $k_0=1$,
but can equally well be generalized to an arbitrary $k_0$. This means that
the constants $\gamma$ and $b$ are interdependent through the relation 
\begin{equation}
\frac{\sum_{k=k_0}^{M}\frac{e^{-bk}}{k^{\gamma-1}}}
{\sum_{k=k_0}^{M}\frac{e^{-bk}}{k^\gamma}} = \frac{M}{N}.
\label{eq:b}
\end{equation}
The third condition is determined by requiring a specific average value for
the size of the largest box $\left<k_{\max}\right>$.
In the direct comparison with a
single dataset, this value is approximated with the actual value of
$k_{\max}$ for the dataset. The calculation of $\left<k_{max}\right>$
is made in two steps: first a value $k_c$ is determined by the condition that
$\sum_{k=k_c}^{M} P(k) = 1/N$. This means that on the average there is
precisely one box in the interval $[k_c,M]$. The second step is
to calculate the average size of a box in the interval $[k_c,M]$, i.e.,
\begin{equation}
\left< k_{\max} \right>=\frac{\sum_{k=k_c}^{\infty} k
P(k)}{\sum_{k=k_c}^{\infty} P(k)}.
\label{eq:eps}
\end{equation}
Thus the three requirements turn into a set of self-consistent equations:
one starts by assuming a certain value for $\gamma$ and then one obtains $b$
by using \eref{eq:b}. Thus the two basic constraints
\eref{eq:basic} yields $P(k)$ from the trial $\gamma$. Next this trial
$P(k)$ is inserted in \eref{eq:eps}. If \eref{eq:eps} is not
satisfied within a predefined precision, we repeat this procedure with a new
trial $\gamma$. In this way, the correct values of $A$, $b$ and $\gamma$ can
be self-consistently determined.

\section{Random Book Transformation}
\label{app:rbt}

Suppose we want the distribution for an $n$th part of a book which has the
word-frequency distribution $P(k)$. The chance that a picked word is part of
the $n$th part is $\frac{1}{n}$ and the chance that it is not is
$\frac{n-1}{n}$. Consequently the RBT gives~\cite{seb09,seb10,baayen01}
\begin{equation*}
P_{\frac{1}{n}}(k)=\frac{\sum_{k'=k}\left(\frac{1}{n-1}\right)^{k}(\frac{n-1
}{n})^{k'}\left (\begin{array}{c} k'\\k \end{array}\right
)P(k')}{1-\sum_{k'=1}(\frac{n-1}{n})^{k'}P(k')} 
\end{equation*}
where $\left (\begin{array}{c} k'\\k \end{array}\right )$ is the
binomial coefficient and the normalization is appropriate because
\begin{eqnarray*}
\sum_{k=1}P_{\frac{1}{n}}(k)
&=& \frac{1}{1-P_{\frac{1}{n}}(0)}\sum_{k' =1}
\left(\frac{n-1}{n}\right)^{k'}P(k')\sum_{k=1}^{k'} \left(\frac{1}{n-1}
\right)^{k}\left (\begin{array}{c} k'\\k \end{array}\right )\\
&=&\frac{1}{1-P_{\frac{1}{n}}(0)}\sum_{k'=1} \left(\frac{n-1}{n}
\right)^{k'} P(k') \left[\left(\frac{n}{n-1}\right)^{k'}-1\right]\\
&=&\frac{1}{1-P_{\frac{1}{n}}(0)} \left[1-P_{\frac{1}{n}}(0)\right]=1
\end{eqnarray*}

The RBT can be analytically obtained in two limiting cases.
These are the equal-address RGF distribution $P(k)\propto \exp(-bk)/k$ and
the limit distribution $P(k)\propto \exp(-bk)/\left[k(k-1)\right]$.
The transformed solution in the first case is given by
\[
P_{\frac{1}{n}}(k)\propto\frac{\exp \{-k\ln [n(e^{b}-1)+1]\}}{k} 
\]
\bigskip
and, since  $(e^{b}-1)=b$ for small $b$, it reduces to
\[
P_{\frac{1}{n}}(k)\propto\frac{\exp [-k\ln (nb+1)]}{k} \propto
\frac{\exp(-knb)}{k}.
\]
This means that the exponential cutoff increases linearly with $n=M/m$, or
in other words, the size dependence is to good approximation given by
$P_m(k)\propto\frac{\exp(-kb_0/m)}{k}$ where $b_0$ is a constant. This
result just reflects the fact that the exponential cuts off the distribution
at the system size $m$. An important consequence is that the functional form
$P(k)\propto 1/k$ is invariant under the RBT. This is a very
special property and $P(k)\propto 1/k$ is presumably the only nontrivial
invariant functional form with a finite value at $k=1$. The typical situation
is that the shape of $P(k)$ becomes less broad under the transformation,
e.g., $P(k)\propto 1/k^\gamma$ with $\gamma >1$ will have an increasing
$\gamma$ with decreasing size. 

The functional form $P(k)\propto \exp(-bk)/[k(k-1)]$ transforms as
\begin{eqnarray}
P_{\frac{1}{n}}(k) &\propto& \frac{1}{k(k-1)} \times \frac{1}{n} \times
\frac{1}{(e^b n-n+1)^{k-1}}\nonumber\\
&\propto& \frac{\exp (-bnk)}{k(k-1)}.
\end{eqnarray}
One notes that the exponent transforms in the same way and has the form
$\exp(-kb_0/m)$. One also notes that the form $P(k)\propto 1/[k(k-1)]$ is
invariant but that it is infinite for $k=1$. The point is that if you start
from $P(k)\propto 1/k^\gamma$ then the transformed $P_m(k)$ approaches the
limit form  $P_\infty(k)\propto 1/[k(k-1)]$. It is also interesting to note
that, for $k$ values not too small, the limiting function is a power law
with exponent $\gamma=2$.

One may then ask what happens if we instead followed the transformation in
the reverse direction towards larger books. Since $P(k)\propto 1/k$ is
invariant under the transformation, it seems likely that it is also the
limiting function in the reverse direction so that $P_0(k)\propto 1/k$. This
suggests that a book approaches this word-frequency distribution in the
limit of infinite size. As seen from the data analysis in \sref{sec:predict},
this expectation seems to have some support in the actual data. One may
perhaps also speculate that since $\gamma=1$ for the upper limit and
$\gamma=2$ for the lower, it should not be a surprise that $\gamma$ values
within $1<\gamma<2$ are often found in real data.

\end{document}